\documentclass[trackchanges]{aastex701}
\usepackage{array}
\usepackage{tabularx}
\usepackage{booktabs}
\usepackage{amsfonts}
\usepackage{amsmath}
\usepackage{subcaption}
\usepackage{multirow}
\usepackage{siunitx}

\usepackage{placeins}

\begin{document}

\title{Short-Term Solar Proton Event Prediction with Energetic Electrons: \\ Comparing Multivariate Interval-Based Ensembles and Neural Transformers}

\author[0000-0003-3196-3822]{Aatiya Ali}
\affiliation{Georgia State University}
\email{aali87@gsu.edu}

\author{Viacheslav Sadykov}
\affiliation{Georgia State University}
\email{vsadykov@gsu.edu}

\author{Berkay Aydin}
\affiliation{Georgia State University}
\email{baydin2@gsu.edu}

\author{Benedict Mervyn}
\affiliation{Georgia State University}
\email{bmervyn1@gsu.edu}

\begin{abstract}
Solar proton events (SPEs) pose substantial radiation risks to spacecraft and human exploration missions. Because sub-relativistic electrons propagate faster than $\sim$10--50 MeV protons and typically arrive at L1 earlier, they provide a potential precursor for short-term SPE forecasting. We compare a Sliding Window Multivariate Time Series Forest (Slim-TSF) and a neural Transformer using electron fluxes from ACE/EPAM and SOHO/EPHIN, together with GOES relativistic electron and soft X-ray (SXR) data, to forecast S1-level SPEs ($\geq$10 pfu at $\geq$10 MeV) from 3-, 6- and 12-hour observation windows at lead times of 15, 30 and 60 minutes, where the lead time is the interval after the end of the observation window within which the proton flux must cross 10 pfu. Each configuration is evaluated on seven walk-forward folds, i.e., chronological training/test splits of the 1997--2020 record, with 30 runs per fold, and performance is reported as the median $\pm$ median absolute deviation (MAD) of the true skill statistic (TSS) and other metrics. Over the nine baseline (no-SXR) configurations, Slim-TSF achieves median TSS of 0.702--0.897, recall of 0.727--1.000, precision of 0.178--0.607 and a false-positive rate (FPR) of 0.009--0.061; the Transformer achieves TSS of 0.586--0.873, recall of 0.675--0.915, precision of 0.185--0.585 and FPR of 0.017--0.042. At 3-hr windows the two are comparable: the Transformer has the higher median TSS at the 15- and 30-min lead times, but Slim-TSF is ahead in 12 of 21 folds. At 6 and 12 hr Slim-TSF leads (0.773--0.897 against 0.586--0.684), most clearly at 12 hr. An ablation shows that the Transformer's decline with window length comes mainly from averaging its encoder output over the window rather than from the changing class balance: using the last time step instead raises its 12-hr TSS from 0.59 to 0.87. SXR integration produces no consistent change for the Transformer and at most a small gain for Slim-TSF, both within the fold-to-fold scatter, and skill depends only weakly on lead time. Both models achieve F1 scores comparable to, and in most comparisons exceeding, those reported by \citet{Posner2007} and \citet{Torres2025}; because the studies differ in evaluation period, event definition and base rate, this comparison is indicative rather than a direct ranking.
\end{abstract}


\section{Introduction}
\setcounter{footnote}{0}
Impacts of solar energetic particle (SEP) events are among the most critical challenges to the longevity of space-based infrastructure and the safety of crewed space exploration. A subclass of SEP events, solar proton events (SPEs), are sudden enhancements of high-energy proton fluxes in the heliosphere. SPEs observed near-Earth can be categorized by the NOAA Solar Radiation Storm scale (S-scale\footnote{\url{https://www.swpc.noaa.gov/noaa-scales-explanation}}), which quantifies the flux of $\geq$10 MeV protons based on particle flux unit (pfu) magnitude thresholds. Events of S3 or greater (with peak fluxes $\geq{10}^{3}$\,pfu) can cause significant radiation damage to electronics and pose acute health risks to astronauts. As the space sector shifts toward sustainable exploration, the success of Artemis and future deep-space missions will rely on operational (and potentially on-spacecraft) forecasting of such events. Short-term forecasts (within $\sim$1 hour of event arrival time) are therefore increasingly essential for operational safety, providing the necessary lead-time to reschedule extravehicular activities or utilize radiation-shielded environments.

\renewcommand{\arraystretch}{0.8}
\begin{deluxetable}{lcccc}
\tablecaption{Scatter-free travel times for particles propagating from the Sun to L1 along a Parker spiral (considered here to have a nominal length of 1.2\,AU). Columns list, respectively, the instrument/particle channel, particle kinetic energy ($E_k$), dimensionless speed ($\beta=v/c$), Sun-L1 travel time ($t$), and the arrival-time advance ($\Delta t$) relative to 10 MeV protons, representing the maximum kinematic warning time available from each channel.
\label{tab:timing}}
\tablehead{
\colhead{Species / channel} & \colhead{$E_k (MeV)$} & \colhead{$\beta$} &
\colhead{t (\textit{mins})} & \colhead{$\Delta t$ wrt. 10 MeV protons (\textit{mins})}}
\startdata
\cutinhead{Electrons}
ACE/EPAM  & 0.045 & 0.394 & 25.3 & 43.6 \\
ACE/EPAM  & 0.062 & 0.452 & 22.1 & 46.8 \\
ACE/EPAM  & 0.175 & 0.667 & 15.0 & 53.9 \\
ACE/EPAM  & 0.312 & 0.784 & 12.7 & 56.2 \\
SOHO/EPHIN & 1.080 & 0.947 & 10.5 & 58.4 \\
SOHO/EPHIN & 3.854 & 0.993 & 10.0 & 58.9 \\
\cutinhead{Protons}
GOES & 10  & 0.145 & 68.9 & \nodata \\
GOES & 30  & 0.247 & 40.4 & 28.5 \\
GOES & 50  & 0.314 & 31.8 & 37.1 \\
GOES & 100 & 0.428 & 23.3 & 45.6 \\ \vspace{-7pt}
\enddata
\tablecomments{The additional propagation time from L1 to GEO ($\sim1.5\times10^{6}$ km) is at most
$\sim0.6$ min and is neglected.}
\end{deluxetable}

On SEP propagation and the physical lead-time limit, the velocity of a particle of kinetic energy $E_k$ and rest energy $E_0$ is:
\begin{equation}
\beta \;=\; \frac{v}{c} \;=\; \sqrt{1 - \left(1 + \frac{E_k}{E_0}\right)^{-2}},
\label{eq:beta}
\end{equation}
\noindent so that the scatter-free travel time along a magnetic field line of path length $s$ is:
\begin{equation}
t \;=\; \frac{s}{\beta c}
\;=\; \frac{s}{c}\left[1 - \left(1 + \frac{E_k}{E_0}\right)^{-2}\right]^{-1/2},
\label{eq:tof}
\end{equation}
\noindent with $E_0 = 0.511$ MeV for electrons and $E_0 = 938.272$ MeV for protons. Table~\ref{tab:timing} evaluates Equation~\ref{eq:tof} for $s = 1.2$ AU, the nominal Parker spiral path length at 1 AU for 400 km s$^{-1}$ solar wind \citep{parker}, across the electron channels adopted as input features and across proton energies at and above the S1 threshold. Relativistic and near-relativistic electrons propagate through the interplanetary medium significantly faster than potentially hazardous $\geq$10 MeV protons, which makes them a viable precursor candidate for SPEs.

Equation~\ref{eq:tof} assumes that every species travels the same path, at zero pitch angle and without scattering. The path itself does not shorten with energy in this range. Charged particles follow the field line through their guiding center as long as their gyroradius is small compared with the scale over which the field changes, and in a 5\,nT field at 1\,AU the largest gyroradius (at a $90^{\circ}$ pitch angle) is $\sim$150--2,900\,km for the 0.045--3.854\,MeV electrons of Table~\ref{tab:timing} and $\sim$0.9--3$\times10^{5}$\,km ($\leq$0.002\,AU) for 10--100\,MeV protons, orders of magnitude below the $\sim$1\,AU curvature scale of the spiral. More energetic particles therefore do not take a shorter route. What departs from Equation~\ref{eq:tof} is the scatter-free assumption: a particle with pitch angle $\alpha$ advances along the field at $v\cos\alpha$, and pitch-angle scattering lengthens the effective path, so onset-time analyses often infer path lengths longer than the nominal spiral \citep{lin1974,Reames1999}. The travel times in Table~\ref{tab:timing} are therefore lower bounds. For a common effective path, a longer path delays slow protons more than fast electrons (an extra 0.3\,AU adds 17\,min for 10\,MeV protons but only 2.5--6\,min for the electron channels), so the arrival-time advances in Table~\ref{tab:timing} are conservative. Throughout this paper, the \textit{lead time} is the forecast horizon: the interval after the end of an observation window within which the $\geq$10\,MeV proton flux must cross 10\,pfu for the window to count as positive (Section~\ref{meth}). It is distinct from the physical electron--proton arrival delay of Table~\ref{tab:timing}, which bounds the lead time an electron-based forecast can support.

The \textit{Relativistic Electron Alert System for Exploration} (REleASE) scheme developed by \citet{Posner2007} utilizes empirical forecasting matrices trained on historical electron data from the \textit{Electron, Proton, Helium INstrument} (EPHIN) on the \textit{Solar and Heliospheric Observatory} (SOHO) to project 30-50 MeV proton fluxes based on the arrival of near-relativistic electrons. REleASE demonstrated that these electrons can arrive at 1 AU between 30 and 90 minutes before the corresponding proton fluxes exceed the 10 pfu threshold (for a baseline S1 SPE). This lead time provides a crucial window for deploying early warnings once an SPE is predicted, and serves as the motivation for this paper. Historically, SEP forecasting was primarily founded upon empirical matrices and linear statistical models \citep{Laurenzaetal2009}. These classical techniques established physical correlations between properties of the solar precursors, such as soft X-ray (SXR) flares and radio bursts, with resulting SEP events \citep{Huangetal2012, Swalwelletal2017}. While such models provided essential early warning capabilities, they often exhibit high false positive (FP; i.e., false alarm) rates due to their limited ability to capture complex, non-linear dependencies inherent in variability of solar activity and heliospheric particle transport, especially when the class imbalance is naturally severe \citep{me, Swalwelletal2017}. This makes machine learning (ML) a prime candidate for forecasting such rare events. A comprehensive review of SEP event prediction models presented in \citet{Whitman2023} demonstrates a strong involvement of ML models emerging as a promising method for capturing these complex dependencies.

In addition, the current literature reveals gaps in the application of advanced architectures to electron-based forecasting. A more recent review of SEP prediction approaches by \citet{Kasapis2026}, which explicitly focuses on ML models, indicates that only a few models address this issue. For instance, \citet{Torres2025} utilized ML methods (such as recurrent neural networks (RNNs) and multilayer perceptrons) to associate relativistic electron and proton fluxes. They found that RNNs predict expected proton fluxes more accurately, albeit with a larger time lag, compared to the earlier approach by \citet{Posner2007}. Other notable efforts include the \textit{Solar Energetic Particle Forecasting with Multi-Task Deep Learning} (SEPNET) model, a multi-task deep learning framework that predicts SEP events by modeling SEP occurrence alongside solar flares and coronal mass ejections (CMEs). To achieve this, SEPNET \citep{sepnet} integrates comprehensive multi-source observations, including historic SXR flux, proton flux, and \textit{Helioseismic and Magnetic Imager} (HMI) Active Region (AR) Patches magnetic field parameters \citep[SHARP,][]{Bobra2014SHARP}. By integrating Long Short-Term Memory (LSTM) and Transformer architectures to model complex temporal dependencies, SEPNET achieves higher event detection rates and superior skill scores in jointly forecasting solar flares, CMEs, and SEPs. The \textit{Multivariate Ensemble of Models for probabilistic forecast of SEPs} (MEMPSEP) model \citep{memsep,chatterjee2} also employs a multi-ensemble approach of convolutional NNs to characterize the SEP environment. Despite these advances, gaps remain in addressing short-term, electron-triggered SPE prediction. For example, isolating the physical relationship between sub-relativistic electron propagation and proton onset from other statistical relationships between energetic protons and solar activity still proves challenging. A key motivation for our approach is that developing advanced warning systems based on in-situ electron measurements can be significantly more practical for deep space mission preparation compared to maintaining continuous, intensive solar monitoring. Our interest in the potentially complex association between energetic electrons and protons for short lead times ($\leq$ 1 hour) has not been widely explored within modern ML architectures \citep{Kasapis2026}.

Transformer-based models utilize multi-head self-attention mechanisms to weigh the importance of different time steps in a given temporal sequence. This allows the model to capture long-range temporal dependencies, such as gradual rises in flux or other signatures that classical models might miss \citep{Vaswani2017}. In parallel, simpler and more interpretable interval-based classifiers present promising alternatives for SEP prediction. A notable example is the Sliding Window Multivariate Time Series Forest (Slim-TSF) \citep{SLIMTSF2024}. Initially developed within the context of solar flare and space weather research, Slim-TSF is designed for global feature explainability and efficiency. It works by applying multi-scale sliding windows to compute descriptive statistics (in this work, mean, standard deviation, and slope) over specific sub-intervals of the given time series. Given that these statistics are acquired during different intervals and vary both in length and lead time, the post-analysis of feature importances can identify which sub-intervals contribute most to model accuracy. While deep learning offers high accuracy, operational space weather forecasting models based on Slim-TSF can potentially be more robust, computationally efficient, and physically interpretable \citep{SLIMTSF2024}.

In this study, we address the aforementioned gaps in SEP event research by explicitly investigating the utility of modern ML architectures for short-term SPE forecasting. We compare how well Slim-TSF and a Transformer capture temporal patterns in the data, evaluate which achieves better overall forecast skill, and examine where each falls on the precision-recall spectrum. We also investigate the performance of these models with time-cumulative importance of related observational data depending on different observation windows (3-, 6-, or 12-hr) and lead times (15-, 30-, or 60-min). To focus on short-term precursors with respect to longer-term statistical associations, we utilize high-energy electron measurements from the \textit{Electron, Proton and Alpha Monitor} (EPAM) on the \textit{Advanced Composition Explorer} (ACE), and SOHO/EPHIN located at the Lagrangian point, L1. These data allow us to evaluate the efficacy of sub-relativistic and near-relativistic electron fluxes as precursors to impending SPEs at the geostationary orbit (GEO). Furthermore, we incorporate NOAA's \textit{Geostationary Operational Environmental Satellite} (GOES) SXR measurements at long (0.1–0.8 nm) and short (0.05–0.4 nm) wavelengths to investigate whether multi-wavelength flare signatures enhance predictive capability and model robustness (especially with respect to potential FP forecasts) between structured statistical windowing and deep neural architectures. We evaluate these models using standard space weather-centric metrics (F1-score, True Skill Statistic (TSS), and Heidke Skill Score (HSS); \citet{Bloomfield2012}).


\section{Data Acquisition and Preprocessing}\label{sec:data}
The long-term dataset constructed for this study spans from August 1997 to March 2020 (solar cycles (SCs) 23-24) and integrates multi-spacecraft measurements of the heliospheric particle environment. We utilize a combination of sub-relativistic and relativistic electron flux observations and SXR flux profiles as model inputs, and high-energy proton fluxes to define the forecast targets. Data were aggregated from the following instruments:

\begin{itemize}
    \item \textbf{ACE/EPAM:} We utilize EPAM data to retrieve sub-relativistic electron fluxes in the 0.045--0.312 MeV range. These data, sourced from the NASA CDAWeb archive\footnote{\url{https://cdaweb.gsfc.nasa.gov/misc/NotesA.html}}, provide critical upstream measurements of electron onsets at the L1 point. Low- to mid-energy electron observations are provided by the Low Energy Foil Spectrometer (LEFS) detectors, a primary component of EPAM aboard ACE \citep{Gold1998}. The EPAM instrument utilizes two distinct telescopes, designated \texttt{LEFS60} and \texttt{LEFS150}, which are oriented at look angles of \ang{60} and \ang{150}, respectively, relative to the spacecraft's spin axis. Both detectors employ an aluminized Parylene foil that suppresses ions with energies $<$ \SI{350}{\kilo\electronvolt} while permitting electrons with energies $>$ \SI{35}{\kilo\electronvolt} to pass through to the solid-state detector arrays. Both \texttt{LEFS60} and \texttt{LEFS150} monitor an identical set of four differential energy bands spanning \SIrange{0.045}{0.062}{\mega\electronvolt}, \SIrange{0.062}{0.103}{\mega\electronvolt}, \SIrange{0.103}{0.175}{\mega\electronvolt}, and \SIrange{0.175}{0.312}{\mega\electronvolt}, and each divides the spacecraft's rotation into four directional sectors. We use the spin-averaged flux of each band. The three lower \texttt{LEFS150} bands (\SIrange{0.045}{0.175}{\mega\electronvolt}) contain only fill values after 19 March 1998 (ACE Science Center, EPAM Level 2 data documentation), when a noise increase in the detector made them unusable, and are not used; the \texttt{LEFS150} \SIrange{0.175}{0.312}{\mega\electronvolt} band is unaffected.

    \item \textbf{SOHO/EPHIN:} To capture higher-energy electron fluxes extending beyond the EPAM thresholds, we supplement the input feature set with electron fluxes from EPHIN, part of the \textit{COmprehensive SupraThermal and Energetic Particle} (COSTEP) suite aboard SOHO \citep{Muller-Mellin1995}. The sensor consists of a stack of six silicon solid-state detectors in a plastic scintillator that serves as an anti-coincidence shield for background rejection. Particle species and energy are determined by the $dE/dx$–$E$ method, in which the energy deposited in the first detector is combined with the total deposition across the stack \citep{Muller-Mellin1995}. Because SOHO is three-axis stabilized and EPHIN is oriented along the nominal IMF direction, 45$^{\circ}$ west of the spacecraft–Sun line, the instrument preferentially samples particles arriving along the nominal Parker spiral \citep{parker}, complementing the spin-sectored, anisotropy-resolving coverage provided by EPAM. During intense events, the outer ring segments of the first two detectors are automatically deactivated, reducing the geometry factor by roughly a factor of 25 to limit dead-time losses. This ring switching requires correction before the count rates can be interpreted as fluxes. We employ the corrected electron flux dataset from the Zenodo repository \citep{zenodo}, which can be specifically utilized to quantify relativistic electron fluxes as a precursor for SPE onset prediction, and use its two differential channels at effective energies of \SI{1.080}{\mega\electronvolt} and \SI{3.854}{\mega\electronvolt}. This series ends on 4 October 2017.

    \item \textbf{GOES:} We leverage the \textit{Energetic Particle Sensor} (EPS) and the \textit{Energetic Proton, Electron, and Alpha Detector} (EPEAD) for both input features and ground-truth validation of model performance. The integral electron channels E1 ($\geq$0.6 MeV; $>$0.8 MeV with revised geometric factors on GOES-13--15), E2 ($\geq$2 MeV) and E3 ($\geq$4 MeV) are included in this study as proxies for the magnetospheric environment. We use all three as delivered, without differencing, so the channels are nested: E1 contains E2, which in turn contains E3. Although the electron population GOES samples at these energies is dominated by the trapped outer-belt population rather than by freshly injected solar electrons, we retain these channels as predictive precursors. For 2000--2002, the GOES-8 EPS electron data were taken from the NCEI 5-min files with the same cross-calibration as the rest of the series. It is worth noting that these channels experience contamination from higher-energy SEPs during intense events \citep{RodriguezATBD2014}. To systematically evaluate how preceding solar flare properties alter forecasting performance (in case of a flare occurring within the observational window), GOES SXR measurements at long (0.1--0.8\,nm) and short (0.05--0.4\,nm) wavelengths are added in the with-SXR model variants. Here, we use the SXR data provided by NOAA National Center for Environmental Information (NCEI\footnote{\url{https://www.ncei.noaa.gov/data/goes-space-environment-monitor/access/avg/}}). To ensure consistency during periods of multiple active GOES satellites, we designated a ``primary'' source/detector each month, choosing the satellite and detector that measured the highest flux that month \citep{me}. Finally, we define binary SPE occurrence labels (forecast targets) based on $\geq$ 10\,MeV GOES proton flux, following the standard SPE threshold defined by NOAA. In this work, $\geq$10 pfu must be maintained for at least 15 minutes to be considered an event, correspondent to three consecutive data points. We note here that while EPHIN and EPAM data are obtained at the L1 point and GOES data are obtained at GEO, SPE measurements are generally consistent at these two vantage points \citep{Ali2026JGRA_GOES_EPHIN} apart from known contamination issues \citep{Posner2007,Ali11415910_MLcontamination}.
\end{itemize}

Table~\ref{tab:inputs} lists the resulting model inputs. All models share a single input set: ten flux channels (four \texttt{LEFS60} bands, one \texttt{LEFS150} band, two EPHIN and three GOES electron channels) and one availability flag per instrument (Section~\ref{sec:cleaning}), 13 inputs in all. The with-SXR variants add the two SXR channels and their flag (16 inputs); the input set does not otherwise differ between models.

\begin{table}[t]
\centering
\footnotesize
\caption{Model inputs, on the common 5-min grid of the proton labels. ``Missing'' gives the fraction of the full record (30 August 1997 to 4 March 2020) without valid data before $\rightarrow$ after the 1-hr forward fill of Section~\ref{sec:cleaning}; the remaining gaps are set to the training mean and marked by the availability flag of the instrument. All model variants use the same inputs, except that the with-SXR variants add the two SXR channels and their flag. Diff.\ = differential, int.\ = integral.}
\label{tab:inputs}
\begin{tabular}{llll}
\toprule
Instrument / channel & Energy or band (type) & Missing & Valid data \\
\midrule
ACE/EPAM LEFS60 ($60^{\circ}$) & 0.045--0.062, 0.062--0.103, & 1.1 $\rightarrow$ 0.8\% & 1997--2020 \\
 & 0.103--0.175, 0.175--0.312 MeV (diff.) & & \\
ACE/EPAM LEFS150 ($150^{\circ}$) & 0.175--0.312 MeV (diff.) & 1.1 $\rightarrow$ 0.8\% & 1997--2020 \\
SOHO/EPHIN E300, E1300 & 1.080, 3.854 MeV effective (diff.) & 21--24 $\rightarrow$ 18\% & 1997--Oct 2017 \\
GOES EPS/EPEAD E1, E2, E3 & $\geq$0.6$^{a}$, $\geq$2, $\geq$4 MeV (int.) & 1.4--1.7 $\rightarrow$ 1.0--1.1\% & 1997--2019 \\
GOES XRS long, short$^{b}$ & 0.1--0.8, 0.05--0.4 nm (irradiance) & 17--27 $\rightarrow$ 1.6--3.7\% & 1997--2020 \\
Availability flags & EPAM, EPHIN, GOES $e^-$, SXR$^{b}$ (binary) & -- & -- \\
\midrule
\multicolumn{4}{l}{\textit{Not used}} \\
ACE/EPAM LEFS150 ($150^{\circ}$) & 0.045--0.062, 0.062--0.103, & 97.7 $\rightarrow$ 97.6\% & 1997--Mar 1998 \\
 & 0.103--0.175 MeV (diff.) & & \\
\bottomrule
\end{tabular}

\vspace{4pt}
\raggedright
$^{a}$\,$>$0.8 MeV, with revised geometric factors, for GOES-13--15.
$^{b}$\,With-SXR variants only. Baseline: 10 flux channels and 3 flags (13 inputs); with SXR: 12 channels and 4 flags (16 inputs).
\end{table}

The two contributions to the offset between the L1 electron measurements and the GOES proton measurement at GEO include spacecraft location and the interval between proton onset and crossing of the 10 pfu threshold defining our target. First, the spacecraft separation itself is minor (with respect to the Earth-Sun line), $1.5\times10^{6}$ km between L1 and the near-Earth environment, adding $<$1 minute of transit time for 10 MeV protons and is therefore negligible in this work. Any EPHIN-to-GOES lag is therefore governed almost entirely by the velocity difference between relativistic electrons and $\geq$10 MeV protons, and only marginally by instrument location. Second, because we exclude event-in-progress windows (Section~\ref{meth}), every positive sample requires the model to anticipate a threshold crossing rather than to recognize an ongoing event, so the available warning time is the dispersion term \textit{plus} the rise time from onset to threshold crossing. Gradual CME-shock-driven SEP events often exhibit extended intensity profiles, whereas impulsive events with favorable magnetic connectivity can produce rapid rises following onset \citep{Reames2013,kahler2001}. Because the fastest particles arrive first through velocity dispersion, high-energy channels provide the earliest kinematic information \citep{lin1974,reames2004}.


\subsection{Data Cleaning}\label{sec:cleaning}
We implemented a standard cleaning pipeline that addresses instrument artifacts and data gaps. (i) Outlier filtering: electron and SXR fluxes are non-negative by construction, so every negative entry is a fill or sentinel value. We mask all negative values as missing in the proton, electron and SXR fluxes, which captures the -1$\cdot$${10}^{31}$-style fills as well as the -999.9, -9999 and -32768 sentinels used by individual products; zero flux values in the input channels are treated as missing as well. (ii) Gap handling: each input channel is forward-filled over gaps of at most 1 hr (12 steps). Longer gaps remain missing and are set to the training mean after standardization, and for each instrument (EPAM, EPHIN, GOES electrons, and GOES SXR when used) a binary availability channel marks the steps at which all of its channels hold a measured or forward-filled value, so the models can distinguish an outage from quiet flux. Because forward filling propagates only earlier values into later timestamps, it cannot transport post-onset information backwards into an observation window. After filling, 0.8\% of the EPAM steps, 1.0--1.1\% of the GOES electron steps, 1.6--3.7\% of the SXR steps and 18\% of the EPHIN steps remain missing, the last mostly because the EPHIN series ends in October 2017 (Table~\ref{tab:inputs}; per fold in Table~\ref{tab:fold_coverage}). (iii) Normalization and scaling: we standardize each channel (subtraction of the mean and division by the standard deviation), with each walk-forward fold’s training partition (folds are defined in Section \ref{meth}). This removes differences in units and absolute level between channels \citep{Visscher2015,Torres2025}. Standardization is an affine transformation, however, and does not change the shape of the heavy-tailed flux distributions: values during transient events (flares for SXR fluxes and particle events for electron fluxes) remain far larger in standardized units than quiet-time variations, and can therefore still dominate the input. The two architectures are affected differently. Slim-TSF is unaffected, because the mean, standard deviation and slope of each sub-interval are each shifted and/or rescaled by a positive factor, so the forest's splits, predictions and feature importances are unchanged. The Transformer has no such invariance, and large event-time values may dominate its input projection, attention and gradients. Although particle flux often follows power-law distributions, both architectures extract their signal from the temporal structure of the flux rise (the onset edge) rather than from the absolute dynamic range. Log-transforms appeared to compress this onset signal and would obscure the rapid rise that distinguishes impending events from the background, so we opted to not apply this transformation.

After processing, the dataset comprises 2,300,595 nominal five-minute cadence time steps spanning 30 August 1997 to 4 March 2020, encompassing the complete labeled record without filtering by event magnitude or characteristics. Of these, 14,264 (0.620\%) have no valid proton measurement, so their label is unknown, and 60,774 (2.642\%) lie inside an SPE and carry a positive label. The labels already encode the event definition: an event starts once $\geq$10 pfu is reached at three consecutive points and ends after six consecutive points below the threshold, so flux oscillations right at the threshold do not fragment a single physical SPE and no further merging is applied. This produces 388 contiguous positive runs, of which 212 begin directly after valid quiet data and are therefore true event onsets; the other runs restart directly after a gap in the proton record and are treated as continuations.


\section{Methodology} \label{meth}
\begin{deluxetable}{lcc}
\tablecaption{Comparison of model architectures and implementation \label{tab:structure_comp}}
\tablehead{
\colhead{Feature} & \colhead{Slim-TSF} & \colhead{Transformer}}
\startdata
Architecture &
\parbox{2.3in}{\vspace{4pt}Early-fused random forest (RF) classifier from multivariate time series, using features computed over different multi-scale sliding windows.\vspace{4pt}} &
\parbox{2.3in}{\vspace{4pt}Custom encoder-only Transformer with 2 layers, 4 attention heads, and 64 hidden dimensions.\vspace{6pt}} \\
Input Processing &
\parbox{2.3in}{\vspace{4pt}Relies on global statistical features (mean, standard deviation, slope) over sliding windows.\vspace{4pt}} &
\parbox{2.3in}{\vspace{4pt}Processes the raw, standardized time-series sequence directly through a linear input projection layer.\vspace{6pt}} \\
Feature Management &
\parbox{2.3in}{\vspace{4pt}Statistical features arranged before training; feature ranks computed using block permutation.\vspace{4pt}} &
\parbox{2.3in}{\vspace{4pt}Retains the full sequence with no internal selection; feature ranks computed using block permutation.\vspace{6pt}} \\
Computational Cost &
\parbox{2.3in}{\vspace{4pt}RF on window statistics; multi-core CPU.\vspace{4pt}} &
\parbox{2.3in}{\vspace{4pt}Self-attention memory scales quadratically with window length; GPU.\vspace{6pt}} \\
Hyperparameter Tuning &
\parbox{2.3in}{\vspace{4pt}Halving search done over \texttt{n\_estimators}, \texttt{max\_depth} and \texttt{class\_weight}, based on TSS.\vspace{4pt}} &
\parbox{2.3in}{\vspace{4pt}Fixed configuration held constant across all folds, threshold optimized for TSS.\vspace{6pt}} \\
Imbalance Mitigation &
\parbox{2.3in}{\vspace{4pt}Model class weight options optimized during hyperparameter tuning.\vspace{4pt}} &
\parbox{2.3in}{\vspace{4pt}Enforced via class-weighted cross-entropy at a fixed 1:5 penalty ratio (see Eq.~\ref{eq:loss}).\vspace{6pt}} \\
Optimization Target &
\parbox{2.3in}{\vspace{4pt}Maximizes TSS by balancing precision and recall during parameter sweeps.\vspace{4pt}} &
\parbox{2.3in}{\vspace{4pt}Minimizes class-weighted cross-entropy loss for ten epochs; no early stopping or checkpoint selection (the final-epoch weights are used).\vspace{6pt}}
\enddata
\end{deluxetable}

\begin{figure}
    \centering
    \begin{subfigure}{0.49\linewidth}
        \centering
        \includegraphics[width=\linewidth]{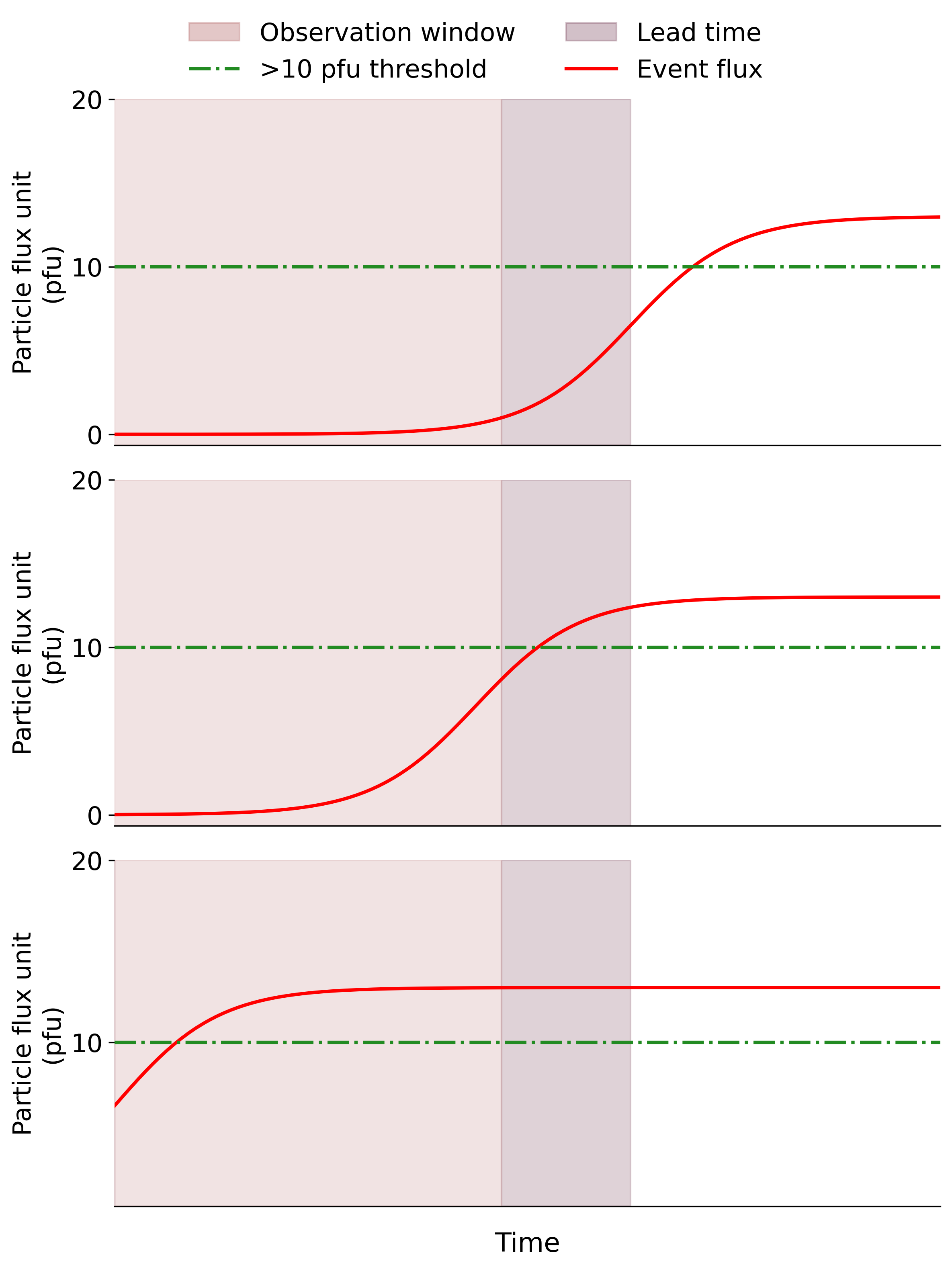}
        \caption{Schematic}
    \end{subfigure}
    \hfill
    \begin{subfigure}{0.49\linewidth}
        \centering
        \includegraphics[width=\linewidth]{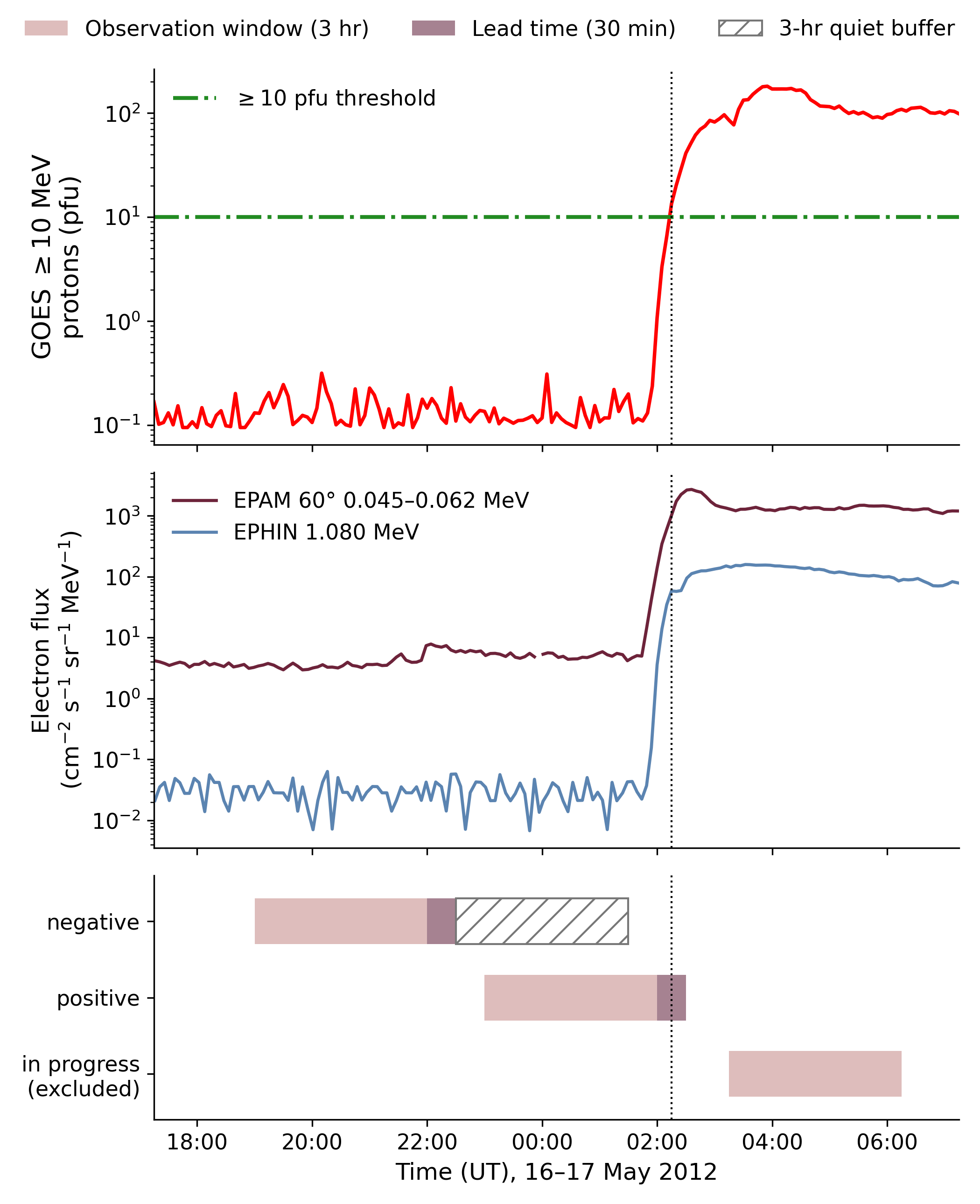}
        \caption{17 May 2012}
    \end{subfigure}
    \caption{Window labeling. (a) Schematic of a negative sample (top), a positive sample (middle) and an event in progress, which is excluded (bottom). (b) The same rule applied to the SPE of 17 May 2012 for the 3-hr observation window and 30-min lead time: GOES $\geq$10 MeV proton flux (top), the EPAM 60$^{\circ}$ 0.045--0.062 MeV and EPHIN 1.080 MeV electron fluxes used as inputs (middle), and one example of each sample type (bottom). The dotted line marks the 10 pfu crossing; the electron fluxes rise about half an hour before it.}
    \label{fig:samples}
\end{figure}

\textbf{Samples.} Each forecast is issued at a time $t$ from an observation window covering the preceding 3, 6 or 12 hr (36, 72 or 144 five-minute steps), and the lead time is the following 15, 30 or 60 min. Figure~\ref{fig:samples} illustrates the labeling with a schematic (a) and a real event (b). (i) A window is \textit{positive} if the $\geq$10 MeV proton flux crosses 10 pfu (an event onset) inside the lead time and no event is in progress during the observation window, providing a clear onset signal for the model to associate with the preceding electron flux. (ii) It is \textit{negative} if the proton flux stays below the threshold throughout the observation window, the lead time and a 3-hr buffer after the lead time, so that windows that precede an onset by only a little more than the lead time are not counted as negatives (near-miss exclusion). (iii) Windows in which an event is already in progress during the observation window are excluded from training and test alike, as the objective is to predict event onset rather than classify ongoing SPEs, and any window touching a gap in the proton record is dropped. Positive windows are issued at every 5-min step, so one onset yields up to 3, 6 or 12 overlapping positive windows for the 15-, 30- and 60-min leads, whereas negative windows are non-overlapping (stride equal to the observation window), which reduces the class imbalance (Table~\ref{tab:samples}). The imbalance nevertheless remains severe, from 10:1 to 142:1 depending on the window and lead time. Overlapping positives are strongly dependent, so the effective sample size is set by the number of distinct SPEs rather than by the number of windows: between 105 (12-hr windows) and 134 (3-hr windows) of the 212 onsets supply at least one positive window.

\textbf{Training and evaluation design.} Each of the nine window--lead time configurations is evaluated with seven embargoed walk-forward \textit{folds}. A fold is a data split, not a model version: the chronologically ordered samples are cut into seven blocks, each containing 25\% of the samples and advancing in steps of 12.5\%, and every block is split into a training partition (about 70\% of the block) and a later test partition, separated by an embargo of $T+H$ rows (the observation window plus the lead time) so that no training window overlaps a test window in time. The split is moved away from 70\% where necessary so that each side contains at least two distinct events. Within each fold, training strictly precedes testing and the two partitions share no samples; across folds the blocks overlap by half, so the per-fold scores are not mutually independent. In five of the seven folds of every configuration, both models are fitted on the first $\sim$80\% of the training partition, and the remaining $\sim$20\% (the validation slice, separated from the fitting data by a $T+H$ embargo) is used only to choose the decision threshold; this split is identical for Slim-TSF and the Transformer. In the third and fourth fold of every configuration the training partition cannot be split into a fitting part with at least two events and a validation slice with at least one, so the whole training partition is used for fitting and a static threshold of 0.30 is applied (540 of the 1,890 fold-run results of each model variant). The splits are held fixed, and every fold is trained and tested 30 times (\textit{runs}) with different random seeds: for Slim-TSF the seed governs the bootstrap resampling and feature subsampling of the RF, and for the Transformer the initialization of its weights and the ordering of the training batches. Each model variant therefore comprises $9 \times 7 \times 30 = 1{,}890$ separately trained models (7,560 for the four variants), each scored on the test partition of its fold. All metrics are reported as the median over the 210 fold-run results of a configuration, with the median absolute deviation, MAD $= \mathrm{median}(|x_i - \mathrm{median}(x)|)$, unscaled (no 1.4826 factor), as the spread. The MAD combines run-to-run variation at fixed splits with the systematic differences between folds, and the second term dominates: the standard deviation of the TSS across seeds within a fold has a median of 0.08 for the Transformer and 0.01 for Slim-TSF, against 0.20 and 0.25 for the spread of the fold-mean TSS. Because the folds overlap, the MAD measures split sensitivity rather than a confidence interval. We treat TSS as the primary metric because, unlike HSS and F1, it does not depend on the ratio of positive to negative windows in the test fold, which varies strongly between folds and rises with window length and lead time.

\textbf{Slim-TSF.} Slim-TSF \citep{SLIMTSF2024} compresses each observation window into statistical aggregates (the mean, standard deviation and slope over multiple sub-intervals), on which an internal RF classifier \citep{Breiman2001,fisher2019} is trained. We utilize \texttt{HalvingGridSearchCV} (successive halving with factor 3 and two-fold internal cross-validation within the fitting portion, scored on TSS) over the grid \texttt{n\_estimators}: [100, 300, 500, 1000], \texttt{max\_depth}: [2, 5, 10, 20, None], \texttt{class\_weight}: [\texttt{balanced}, \texttt{balanced\_subsample}], executed once per fold in the first run; the remaining runs reuse the selected parameters \citep{pedregosa, jamieson}. The with-SXR variant reuses the hyperparameters selected for the corresponding baseline fold, so that the only difference between the two Slim-TSF variants is the input set, while its forests are refitted on the 16 inputs.

\textbf{Transformer.} The Transformer uses an encoder-only architecture with 2 layers, 4 attention heads, and a hidden dimension of 64, held fixed across all windows, lead times and folds. Each input time step (13 or 16 inputs) is projected linearly to the hidden dimension, a sinusoidal positional encoding is added so that self-attention is not blind to temporal order, and the encoder output is averaged over time before a linear two-class output head. It is implemented within the Hugging Face ecosystem, utilizing its \texttt{Trainer} API\footnote{\url{https://huggingface.co/docs/transformers/en/main_classes/trainer}}, and trained with a batch size of 256, the Trainer's default \texttt{AdamW} optimizer, learning rate 5$\cdot$${10}^{-5}$ and linear decay \citep{wolf2020,loshchilov2019}. Because an exhaustive search was computationally prohibitive, the training budget was fixed at ten epochs for all configurations, by which the validation loss has flattened at 3 and 6 hr and decreases only slowly at 12 hr (Appendix~\ref{app:convergence}). We did not use the validation slice for early stopping or checkpoint selection, for three reasons: the slices are small and hold only a few distinct events, so a validation-loss criterion would be dominated by one or two events; two of the seven folds have no validation slice at all; and a fixed budget keeps the training procedure identical across folds and configurations. The cost of this choice is assessed in Appendix~\ref{app:convergence}. To handle the sparsity of SPE events, the model is trained with a class-weighted cross-entropy loss, implemented in PyTorch \citep{paszke2019pytorch}:
\begin{equation}
    \mathcal{L} = -\frac{1}{N} \sum_{i=1}^{N}
        w_{y_i} \, \log p\!\left(y_i \mid \mathbf{x}_i\right),
    \label{eq:loss}
\end{equation}

\noindent
where $N$ is the number of training samples, $y_i \in \{0, 1\}$ is the
ground-truth label for sample $i$, and
\begin{equation}
    p\!\left(y_i \mid \mathbf{x}_i\right) =
        \frac{\exp\!\left(z_{i,y_i}\right)}
             {\sum_{c \in \{0,1\}} \exp\!\left(z_{i,c}\right)}
    \label{eq:softmax}
\end{equation}
is the softmax probability the model assigns to the true class, with $z_{i,c}$ the logit of class $c$. The weights are fixed at $w_0 = 1$ and $w_1 = 5$, a penalty ratio that recovers high recall without severely degrading precision, independent of each fold's class distribution.

\textbf{Decision threshold.} Both models output a positive (class-1) score: the output of Equation~\ref{eq:softmax} for the Transformer, and the forest vote fraction for Slim-TSF. A window is declared positive when that score reaches a decision threshold, chosen separately for every fold and model by maximizing the TSS over a grid of values from 0.01 to 0.99 (step 0.01) on the validation slice, with ties resolved by averaging the tied thresholds; the static 0.30 is used in the two folds without a validation slice. Over all results the median threshold is 0.30 in every variant, with a range of 0.01--0.92. The same rule is applied to both architectures, so the confusion matrices in Section~\ref{sec:cm} are directly comparable, and the threshold is also the operating point at which the permutation importances of Section~\ref{sec:fi} are scored.


\subsection{Feature Importance} \label{sec:fi}
Given that Slim-TSF is based on internal RFs, cumulative Gini feature importance can be computed for each input channel. In contrast, the Transformer retains the full high-dimensional feature space and has no comparable ranking scheme: its attention weights describe how the model distributes representational context, and are not a direct measure of predictive contribution \citep{Vaswani2017}. Because the two architectures compute incomparable internal quantities, we do not use either model's native diagnostic. Instead, we apply block permutation importance, a single model-agnostic scheme, to both models; a similar input-shuffling strategy has been used to identify the solar-wind drivers of a neural-network forecast of geomagnetic indices \citep{Blunier2021}. Let $\mathbf{X}_{k} \in \mathbb{R}^{M_{k} \times C \times L}$ be the input dataset of test fold \textit{$k$}, comprising \textit{$M_{k}$} samples, \textit{$C$} input channels and \textit{$L$} timesteps, and let $\hat{f}_{k}$ be the model trained on the training partition of fold \textit{k}. The fold's TSS before interfering with anything is:

\begin{equation}
    s_{k} = \mathrm{TSS}\!\left(\mathbf{y}_{k},\,
    \hat{f}_{k}(\mathbf{X}_{k})\right),
    \label{eq:pi-baseline}
\end{equation}

\noindent where \textit{$\mathbf{y}_{k}$} are the corresponding labels. For a channel group $g \subseteq \{1,\dots,C\}$ and a permutation \textit{$\pi$} of the sample indices, we construct a perturbed tensor $\mathbf{X}_{k}^{(g,\pi)}$ where \citep{Breiman2001,fisher2019}:

\begin{equation}
    \bigl[\mathbf{X}_{k}^{(g,\pi)}\bigr]_{n,c,\ell} =
    \begin{cases}
        [\mathbf{X}_{k}]_{\pi(n),\,c,\,\ell}, & c \in g,\\
        [\mathbf{X}_{k}]_{n,\,c,\,\ell},      & c \notin g,
    \end{cases}
    \label{eq:pi-permute}
\end{equation}

\noindent so that sample \textit{$n$} receives the full time series of the random sample \textit{$\pi(n)$} in every channel of \textit{$g$}, while the remaining channels are untouched. Permuting whole time series rather than individual time steps preserves the temporal autocorrelation within each channel, so the procedure removes \textit{only} the association between the group and the target. A single shared \textit{$\pi$} is applied across all members of \textit{$g$}, and the importance of the group is the mean loss of skill over \textit{$R$} independent permutations:

\begin{equation}
    \mathrm{PI}_{k}(g) = s_{k} - \frac{1}{R}\sum_{r=1}^{R}
    \mathrm{TSS}\!\left(\mathbf{y}_{k},\,
    \hat{f}_{k}\!\left(\mathbf{X}_{k}^{(g,\pi^{(r)})}\right)\right),
    \label{eq:pi-group}
\end{equation}

\noindent with $g = \{c\}$ recovering the single-channel case, which is the only case evaluated here. Expressing the loss as a fraction of the unperturbed skill and taking the median over folds gives the score (folds with $s_{k}=0$, for which the fraction is undefined, are omitted):

\begin{equation}
    \mathrm{Imp}(g) =
    \operatorname*{median}_{k=1,\dots,K}\left[\frac{\mathrm{PI}_{k}(g)}{s_{k}}\right].
    \label{eq:pi-final}
\end{equation}

\noindent The scores are computed for the first seed on each of the 63 folds of every model variant, with $R = 5$. They are relative degradations and \textit{not} constrained to sum to unity: if several channels are individually indispensable the total exceeds 1, whereas if the input carries redundancy the sum falls below 1, because any single channel can be removed at little cost. Channels carrying no usable information can return small negative values, which should be read as zero within permutation noise.


\section{Results}\label{sec:results}
The baseline variants exclude SXR from the inputs, and every comparison is made on identical folds and seeds. Because each comparison involves two models, we use comparatives (``higher'', ``lower''); when more than two variants are compared, as in the bold entries of Tables~\ref{tab:skill_part1}--\ref{tab:skill_part2}, we use superlatives.

\begin{table*}[t]
\centering
\small
\setlength{\tabcolsep}{5pt}
\caption{Forecast skill metrics (TSS, HSS, F1) for Slim-TSF and Transformer, reported as median\,$\pm$\,MAD (Median Absolute Deviation) over 30 runs and the seven walk-forward validation folds of each configuration (210 fold-run pairs). T = Transformer, S = Slim-TSF, +X = with SXR input. A score of 1 is perfect. Best median in each row in bold.}
\label{tab:skill_part1}
\begin{tabular}{cc cccc}
\toprule
Lookback (hr) & Lead (min) & T & T+X & S & S+X \\
\midrule
\multicolumn{6}{c}{\textbf{TSS}} \\
\midrule
\multirow{3}{*}{3} & 15 & 0.797$\pm$0.195 & 0.792$\pm$0.202 & 0.702$\pm$0.212 & \textbf{0.809$\pm$0.189} \\
& 30 & 0.873$\pm$0.121 & 0.787$\pm$0.206 & 0.785$\pm$0.213 & \textbf{0.884$\pm$0.115} \\
& 60 & 0.830$\pm$0.162 & \textbf{0.845$\pm$0.148} & 0.837$\pm$0.160 & 0.725$\pm$0.225 \\
\cmidrule(lr){1-6}
\multirow{3}{*}{6} & 15 & 0.684$\pm$0.185 & 0.500$\pm$0.113 & \textbf{0.773$\pm$0.218} & \textbf{0.773$\pm$0.217} \\
& 30 & 0.677$\pm$0.167 & 0.735$\pm$0.236 & 0.776$\pm$0.216 & \textbf{0.879$\pm$0.118} \\
& 60 & 0.624$\pm$0.162 & 0.672$\pm$0.165 & \textbf{0.791$\pm$0.206} & 0.765$\pm$0.206 \\
\cmidrule(lr){1-6}
\multirow{3}{*}{12} & 15 & 0.641$\pm$0.184 & 0.571$\pm$0.180 & 0.897$\pm$0.103 & \textbf{0.945$\pm$0.055} \\
& 30 & 0.586$\pm$0.201 & 0.627$\pm$0.238 & 0.880$\pm$0.113 & \textbf{0.957$\pm$0.043} \\
& 60 & 0.618$\pm$0.162 & 0.599$\pm$0.167 & 0.789$\pm$0.114 & \textbf{0.790$\pm$0.159} \\
\midrule
\multicolumn{6}{c}{\textbf{HSS}} \\
\midrule
\multirow{3}{*}{3} & 15 & \textbf{0.295$\pm$0.132} & 0.242$\pm$0.092 & 0.264$\pm$0.264 & 0.286$\pm$0.240 \\
& 30 & 0.378$\pm$0.110 & 0.383$\pm$0.129 & 0.400$\pm$0.145 & \textbf{0.527$\pm$0.174} \\
& 60 & 0.485$\pm$0.136 & 0.502$\pm$0.137 & \textbf{0.539$\pm$0.149} & 0.485$\pm$0.151 \\
\cmidrule(lr){1-6}
\multirow{3}{*}{6} & 15 & 0.304$\pm$0.164 & 0.259$\pm$0.151 & 0.303$\pm$0.082 & \textbf{0.312$\pm$0.071} \\
& 30 & 0.462$\pm$0.172 & 0.441$\pm$0.151 & 0.556$\pm$0.178 & \textbf{0.612$\pm$0.168} \\
& 60 & 0.532$\pm$0.153 & 0.509$\pm$0.103 & 0.540$\pm$0.211 & \textbf{0.714$\pm$0.157} \\
\cmidrule(lr){1-6}
\multirow{3}{*}{12} & 15 & 0.374$\pm$0.175 & 0.336$\pm$0.184 & 0.425$\pm$0.089 & \textbf{0.429$\pm$0.105} \\
& 30 & 0.518$\pm$0.204 & 0.453$\pm$0.169 & 0.682$\pm$0.138 & \textbf{0.771$\pm$0.094} \\
& 60 & 0.554$\pm$0.164 & 0.554$\pm$0.138 & 0.610$\pm$0.181 & \textbf{0.721$\pm$0.161} \\
\midrule
\multicolumn{6}{c}{\textbf{F1}} \\
\midrule
\multirow{3}{*}{3} & 15 & \textbf{0.304$\pm$0.130} & 0.249$\pm$0.080 & 0.272$\pm$0.266 & 0.286$\pm$0.243 \\
& 30 & 0.391$\pm$0.108 & 0.398$\pm$0.117 & 0.411$\pm$0.133 & \textbf{0.537$\pm$0.168} \\
& 60 & 0.503$\pm$0.109 & 0.521$\pm$0.108 & \textbf{0.558$\pm$0.140} & 0.504$\pm$0.104 \\
\cmidrule(lr){1-6}
\multirow{3}{*}{6} & 15 & 0.317$\pm$0.162 & 0.288$\pm$0.156 & 0.316$\pm$0.061 & \textbf{0.324$\pm$0.055} \\
& 30 & 0.480$\pm$0.156 & 0.455$\pm$0.146 & 0.558$\pm$0.153 & \textbf{0.615$\pm$0.172} \\
& 60 & 0.563$\pm$0.133 & 0.561$\pm$0.099 & 0.544$\pm$0.194 & \textbf{0.716$\pm$0.157} \\
\cmidrule(lr){1-6}
\multirow{3}{*}{12} & 15 & 0.400$\pm$0.159 & 0.361$\pm$0.172 & 0.436$\pm$0.092 & \textbf{0.444$\pm$0.111} \\
& 30 & 0.537$\pm$0.159 & 0.496$\pm$0.159 & 0.701$\pm$0.126 & \textbf{0.774$\pm$0.099} \\
& 60 & 0.595$\pm$0.130 & 0.614$\pm$0.121 & 0.663$\pm$0.150 & \textbf{0.751$\pm$0.101} \\
\bottomrule
\end{tabular}
\tablecomments{Baseline Transformer attains the higher median TSS in all three 3-hr configurations (0.797--0.873 against 0.702--0.837), a lead that rests on three folds in which Slim-TSF misses nearly all events (see text), and Slim-TSF in all six 6- and 12-hr configurations (0.773--0.897 against 0.586--0.684). Over all nine configurations Slim-TSF has the higher HSS in 7 and the higher F1 in 6. Observation window has a larger effect on the median TSS than lead time (15/30/60 min) in both models, and the gap between the architectures changes sign between 3 and 6 hr and widens with window length in favour of Slim-TSF.}
\end{table*}

\begin{table*}[t]
\centering
\small
\setlength{\tabcolsep}{5pt}
\caption{Forecast detection metrics (Precision, TPR, FPR) for Slim-TSF and Transformer, reported as median\,$\pm$\,MAD (Median Absolute Deviation) over 30 runs and the seven walk-forward validation folds of each configuration (210 fold-run pairs). T = Transformer, S = Slim-TSF, +X = with SXR input. A score of 1 is perfect for Precision and TPR; for FPR lower is better. Best median in each row in bold.}
\label{tab:skill_part2}
\begin{tabular}{cc cccc}
\toprule
Lookback (hr) & Lead (min) & T & T+X & S & S+X \\
\midrule
\multicolumn{6}{c}{\textbf{Precision}} \\
\midrule
\multirow{3}{*}{3} & 15 & 0.185$\pm$0.090 & 0.158$\pm$0.064 & 0.178$\pm$0.178 & \textbf{0.223$\pm$0.164} \\
& 30 & 0.261$\pm$0.091 & 0.267$\pm$0.094 & 0.330$\pm$0.174 & \textbf{0.382$\pm$0.218} \\
& 60 & 0.352$\pm$0.091 & 0.370$\pm$0.092 & 0.421$\pm$0.174 & \textbf{0.429$\pm$0.172} \\
\cmidrule(lr){1-6}
\multirow{3}{*}{6} & 15 & \textbf{0.208$\pm$0.134} & 0.181$\pm$0.110 & 0.205$\pm$0.057 & 0.205$\pm$0.043 \\
& 30 & 0.340$\pm$0.138 & 0.343$\pm$0.127 & 0.426$\pm$0.163 & \textbf{0.444$\pm$0.169} \\
& 60 & 0.456$\pm$0.126 & 0.451$\pm$0.102 & 0.403$\pm$0.096 & \textbf{0.682$\pm$0.118} \\
\cmidrule(lr){1-6}
\multirow{3}{*}{12} & 15 & 0.288$\pm$0.139 & 0.263$\pm$0.171 & 0.333$\pm$0.076 & \textbf{0.353$\pm$0.093} \\
& 30 & 0.495$\pm$0.191 & 0.464$\pm$0.172 & 0.545$\pm$0.087 & \textbf{0.667$\pm$0.167} \\
& 60 & 0.585$\pm$0.155 & 0.580$\pm$0.135 & 0.607$\pm$0.089 & \textbf{0.704$\pm$0.122} \\
\midrule
\multicolumn{6}{c}{\textbf{TPR (recall)}} \\
\midrule
\multirow{3}{*}{3} & 15 & \textbf{0.915$\pm$0.085} & 0.818$\pm$0.182 & 0.727$\pm$0.245 & 0.909$\pm$0.091 \\
& 30 & 0.913$\pm$0.087 & 0.812$\pm$0.188 & 0.914$\pm$0.086 & \textbf{0.943$\pm$0.057} \\
& 60 & 0.903$\pm$0.097 & \textbf{0.904$\pm$0.096} & 0.879$\pm$0.121 & 0.896$\pm$0.104 \\
\cmidrule(lr){1-6}
\multirow{3}{*}{6} & 15 & 0.741$\pm$0.241 & 0.500$\pm$0.100 & \textbf{0.864$\pm$0.136} & 0.856$\pm$0.144 \\
& 30 & 0.833$\pm$0.118 & 0.792$\pm$0.208 & 0.886$\pm$0.114 & \textbf{0.917$\pm$0.083} \\
& 60 & 0.792$\pm$0.142 & 0.754$\pm$0.162 & \textbf{0.900$\pm$0.100} & 0.817$\pm$0.183 \\
\cmidrule(lr){1-6}
\multirow{3}{*}{12} & 15 & 0.708$\pm$0.208 & 0.600$\pm$0.300 & \textbf{1.000$\pm$0.000} & \textbf{1.000$\pm$0.000} \\
& 30 & 0.778$\pm$0.222 & 0.667$\pm$0.238 & 0.900$\pm$0.100 & \textbf{1.000$\pm$0.000} \\
& 60 & 0.675$\pm$0.175 & 0.735$\pm$0.207 & 0.792$\pm$0.170 & \textbf{0.812$\pm$0.188} \\
\midrule
\multicolumn{6}{c}{\textbf{FPR}} \\
\midrule
\multirow{3}{*}{3} & 15 & 0.017$\pm$0.015 & 0.017$\pm$0.015 & 0.010$\pm$0.010 & \textbf{0.010$\pm$0.010} \\
& 30 & 0.020$\pm$0.020 & \textbf{0.016$\pm$0.014} & 0.016$\pm$0.015 & 0.016$\pm$0.015 \\
& 60 & 0.030$\pm$0.029 & 0.023$\pm$0.021 & 0.018$\pm$0.018 & \textbf{0.018$\pm$0.015} \\
\cmidrule(lr){1-6}
\multirow{3}{*}{6} & 15 & 0.026$\pm$0.022 & 0.026$\pm$0.022 & \textbf{0.009$\pm$0.008} & 0.010$\pm$0.007 \\
& 30 & 0.036$\pm$0.024 & 0.029$\pm$0.020 & 0.014$\pm$0.012 & \textbf{0.010$\pm$0.008} \\
& 60 & 0.042$\pm$0.028 & 0.038$\pm$0.029 & 0.020$\pm$0.018 & \textbf{0.019$\pm$0.017} \\
\cmidrule(lr){1-6}
\multirow{3}{*}{12} & 15 & 0.030$\pm$0.026 & 0.031$\pm$0.027 & 0.023$\pm$0.021 & \textbf{0.022$\pm$0.020} \\
& 30 & 0.036$\pm$0.031 & 0.032$\pm$0.028 & 0.010$\pm$0.010 & \textbf{0.008$\pm$0.008} \\
& 60 & 0.039$\pm$0.034 & 0.040$\pm$0.036 & 0.061$\pm$0.056 & \textbf{0.038$\pm$0.035} \\
\bottomrule
\end{tabular}
\tablecomments{Slim-TSF holds the higher median recall in 7 of nine configurations (0.727-1.000 against 0.675-0.915 for the Transformer), the Transformer in the three 3-hr configurations. Precision is split (3 configurations for the Transformer, 6 for Slim-TSF, 0 tie), and Slim-TSF has the lower FPR in 8 of nine (0.009-0.061 against 0.017-0.042); its 12-hr, 60-min configuration is the exception (median FPR 0.061).}
\end{table*}

\subsection{Model Performance}\label{sec:metrics}
\begin{figure} 
\centering  
\includegraphics[width=.99\textwidth]{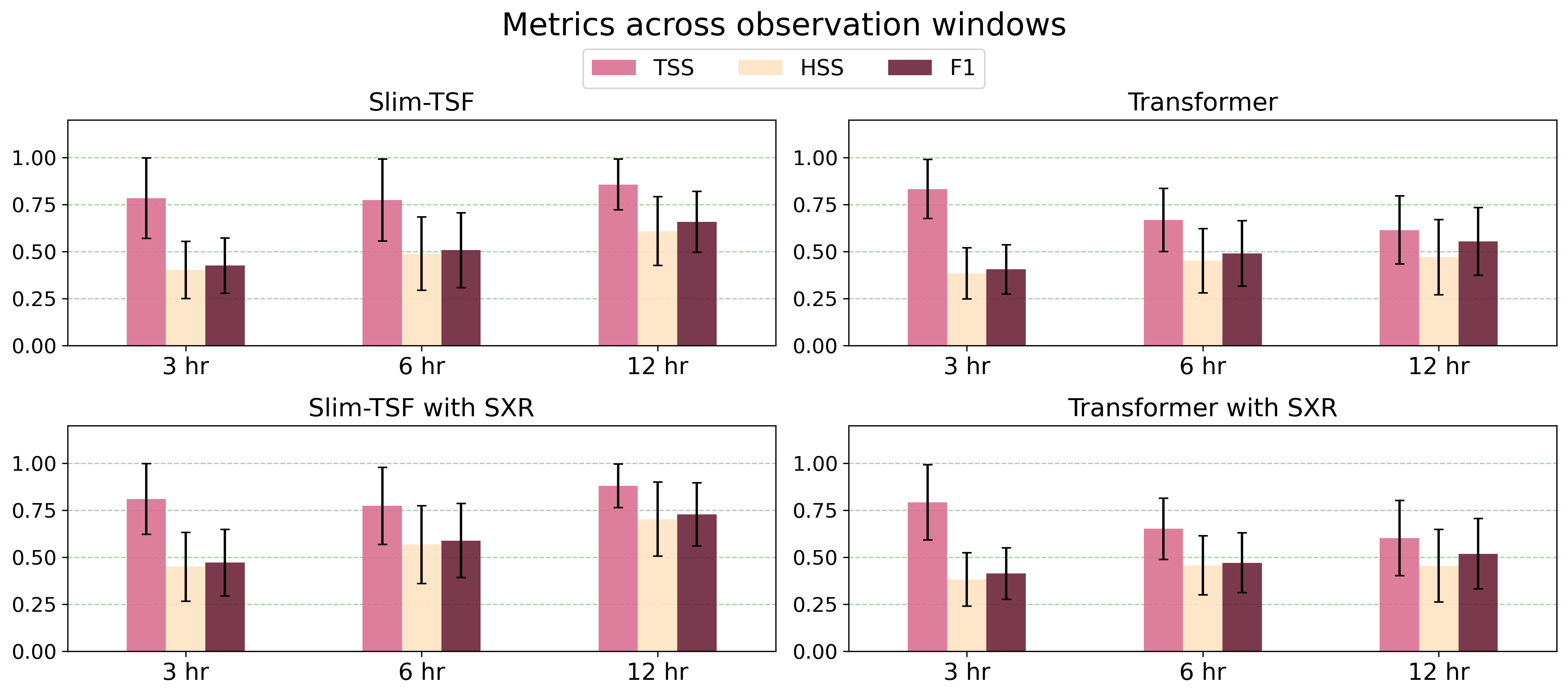}
   \caption{Comparative performance metrics for Slim-TSF and Transformer models across 3-hr, 6-hr, and 12-hr observation periods. Each bar is the median over the 30 runs of each configuration and the three lead times; error bars are the MAD. The top panels illustrate the 13-feature baseline configuration, while the bottom panels show the 15-feature set including GOES SXR data.}
 \label{fig:obs}
\end{figure}
Tables~\ref{tab:skill_part1}--\ref{tab:skill_part2} list the nine configurations for all four variants, and Figure~\ref{fig:obs} summarizes median TSS, HSS and F1 by observation window. With only seven overlapping folds behind each median, and only two distinct SPEs in the test partition of three of them (Table~\ref{tab:fold_detail}), the scatter is large for all skill scores (MAD of 0.103--0.218 for TSS and 0.061--0.266 for HSS and F1 across the nine baseline configurations of either model) and is often comparable to the difference between the two architectures; the counts of configurations in which one model leads, quoted below, are therefore descriptive and not significance tests.

\textbf{Observation window.} Pooled over lead times, folds and runs, the median TSS of the Transformer falls from 0.833 at 3 hr to 0.668 at 6 hr and 0.614 at 12 hr, whereas Slim-TSF is flat from 3 to 6 hr (0.784 and 0.773) and highest at 12 hr (0.857). At 3 hr the Transformer has the higher median TSS at the 15- and 30-min leads (by 0.095 and 0.088) and Slim-TSF at 60 min (by 0.007); at 6 hr Slim-TSF leads by 0.089--0.167 and at 12 hr by 0.171--0.294. The gap exceeds the larger of the two MADs in all three 12-hr configurations but in none of the 3- or 6-hr configurations. Because both models are evaluated on identical folds and seeds, the comparison can also be made fold by fold, averaging each model's TSS over the 30 runs of a fold. At 12 hr Slim-TSF is ahead in 19 of the 21 folds (median TSS difference 0.122) and at 6 hr in 14 of 21 (0.047). At 3 hr it is ahead in 12 of 21 (median +0.012). The Transformer's lead in the pooled 3-hr median rests on three folds in which Slim-TSF misses nearly all test events (fold-mean TSS $\leq$0.19, against 0.67--0.99 for the Transformer); these carry all of the summed 3-hr difference, and without them the two architectures are indistinguishable (mean Transformer-minus-Slim-TSF difference $-$0.002, median $-$0.026). They are Slim-TSF's only complete failures, whereas the lowest fold-mean TSS of the Transformer is 0.33. Slim-TSF is therefore better at 12 hr, ahead at 6 hr within the scatter, and comparable to the Transformer at 3 hr. The 6-hr margin is also of the same order as the gain the Transformer would obtain from a longer training budget (up to 0.08 in mean TSS; Appendix~\ref{app:convergence}), whereas the 12-hr margin is not.

\textbf{Recall and false alarms.} The Transformer is strongest at the shortest window, with median TSS 0.797--0.873, recall 0.903--0.915, FPR 0.017--0.030 and precision 0.185--0.352 at 3 hr. At longer windows its recall falls (0.741--0.833 at 6 hr and 0.675--0.778 at 12 hr) while its FPR rises only slightly (0.026--0.042 and 0.030--0.039), so the loss in TSS comes mostly from missed events. Slim-TSF keeps a high recall at every window (0.727--0.914 at 3 hr, 0.864--0.900 at 6 hr and 0.792--1.000 at 12 hr) at an FPR of 0.010--0.018, 0.009--0.020 and 0.010--0.061 respectively (the largest value belonging to the 12-hr, 60-min configuration). At 12 hr it therefore recovers more events than the Transformer, by 0.12--0.29 in median recall, at a similar precision (0.333--0.607 against 0.288--0.585) and, at two of the three lead times, a lower FPR. At 3 hr the precision of both models is low (0.178--0.421 for Slim-TSF, 0.185--0.352 for the Transformer) because positive windows make up only 0.7--2.7\% of the samples. Comparing the two baseline models over all nine configurations, Slim-TSF has the higher median TSS in seven (Table~\ref{tab:skill_part1}: all six 6- and 12-hr configurations and 3 hr / 60 min), the higher recall and HSS in seven, the higher F1 and precision in six, and the lower FPR in eight.

\textbf{Lead time.} \begin{figure} 
\centering 
\includegraphics[width=.99\textwidth]{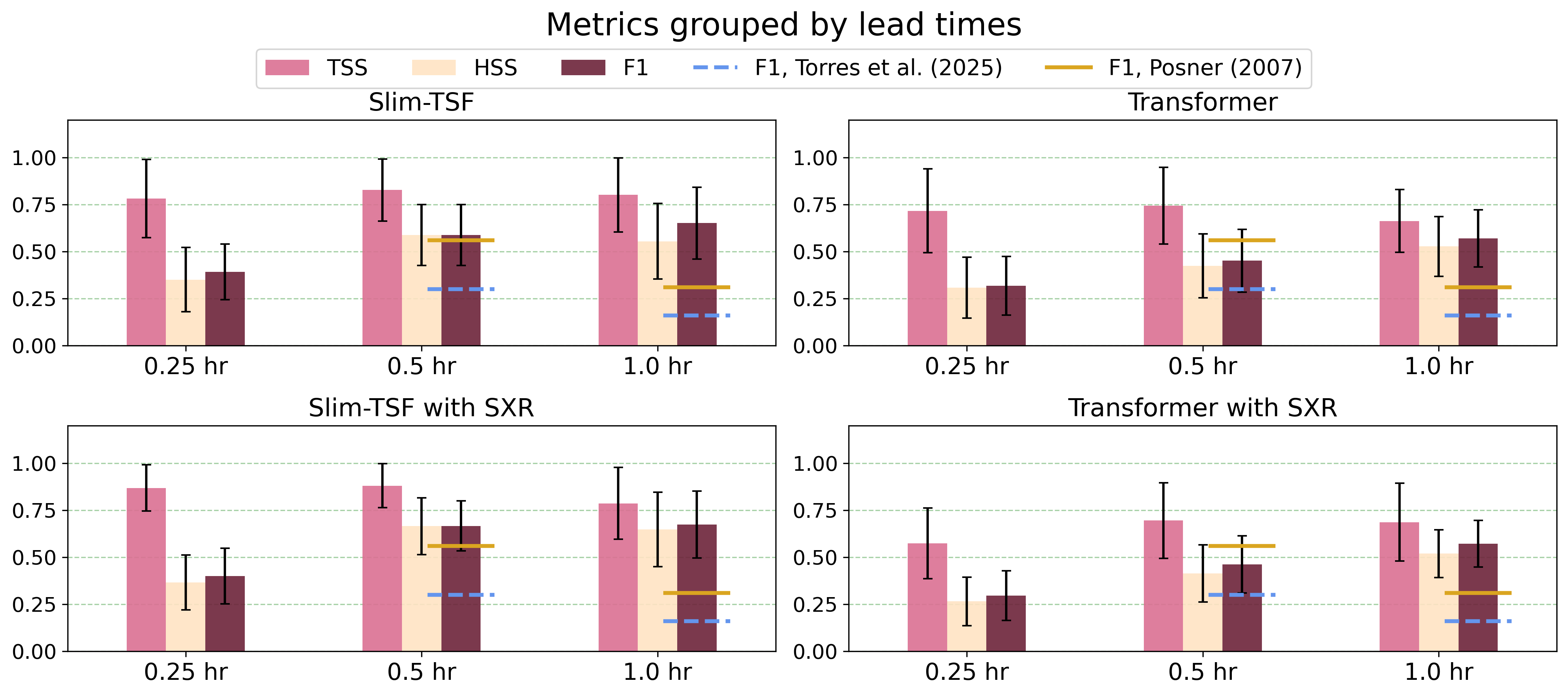}
   \caption{Comparative performance metrics for Slim-TSF and Transformer models across multiple lead times. Each bar is the median over the 30 independent test runs each configuration and the three observation windows; error bars are the MAD. The top panels illustrate the 13-feature baseline configuration, while the bottom panels show the 15-feature set including GOES SXR data.}
 \label{leads}
\end{figure}
Figure~\ref{leads} summarizes the same metrics by lead time. The lead-time dependence is weak. For Slim-TSF the 15-, 30- and 60-min lead times give pooled median TSS of 0.781, 0.827 and 0.801 (MAD 0.208, 0.165 and 0.196); for the Transformer they give 0.717, 0.744 and 0.663 (MAD 0.222, 0.203 and 0.167). The spread across lead times (0.046 for Slim-TSF, 0.082 for the Transformer) is smaller than the scatter within a single lead time, and the 30-min lead is marginally the highest for both models. Within an individual window the medians can differ more (for the Transformer at 3 hr, 0.797, 0.873 and 0.830 at 15, 30 and 60 min), but not monotonically with lead time. Both models therefore appear to have exhausted most of the information available at the end of the observation window, and the operational choice of lead time within this range can be driven by alert-latency and decision-time constraints rather than by skill. HSS and F1, unlike TSS, depend on the positive fraction, and are lowest at 3-hr windows and short lead times (Slim-TSF HSS 0.264--0.539 at 3 hr, against 0.425--0.682 at 12 hr).

\subsection{Class Balance and Pooling}\label{sec:balance}
Because negative windows are non-overlapping while positives are issued at every step, the training set shrinks with window length almost entirely through negatives, and its size and class ratio change together (median training imbalance 61:1, 38:1 and 19:1 at 3, 6 and 12 hr for the 30-min lead). The two models also handle imbalance differently: Slim-TSF's class weights follow each fold's ratio, whereas the Transformer's are fixed at 1:5. A third candidate explanation for the Transformer's decline is its architecture, which averages the encoder output over all time steps, so that the pre-onset rise in the last few steps is diluted four times more at 12 hr (144 steps) than at 3 hr (36 steps). To separate these factors we ran four ablations at the 30-min lead with the 30 seeds of the main experiments on the unchanged test folds (Appendix~\ref{app:ablation}, Table~\ref{tab:ablation}): (a) 3-hr windows with training negatives subsampled to the 12-hr class ratio, (b) 12-hr windows with negatives drawn at the 3-hr stride, (c) the Transformer with per-fold inverse-frequency class weights, and (d) the Transformer classifying from the last encoder state instead of the time average. For the Transformer the class balance explains little of the decline: (a) lowers its 3-hr TSS from 0.873 to 0.830, (b) raises its 12-hr TSS from 0.586 to 0.641, and (c) changes TSS by at most 0.029 at any window. The pooling explains most of it: with the last time step the 12-hr TSS rises to 0.869 (3 hr: 0.916). At 12 hr this is on par with Slim-TSF (0.880), so the architecture ranking at long windows is specific to the mean-pooled Transformer used here. For Slim-TSF the class balance does matter: (a) raises its 3-hr TSS from 0.785 to 0.941 and (b) lowers its 12-hr TSS from 0.880 to 0.770, so its advantage at 12 hr reflects the more balanced 12-hr training set rather than the longer window.

\subsection{Effect of SXR}\label{sec:sxr}
SXR inclusion produces no consistent change in the Transformer: its pooled median TSS shifts by $-$0.040, $-$0.016 and $-$0.012 at 3, 6 and 12 hr, the per-configuration shifts range from $-$0.184 to +0.058 with mixed sign, and fold by fold it changes by $-$0.035 on average (median $-$0.010). For Slim-TSF the median TSS is higher with SXR in six of the nine configurations (shifts of $-$0.112 to +0.107; pooled shifts of +0.026, +0.000 and +0.024); fold by fold the gain averages +0.030 (median +0.002; 62\% of folds improve) and is concentrated at 3 hr with 15- and 30-min leads (mean gains of +0.098 and +0.069, against at most +0.035 elsewhere). No configuration-level shift exceeds the corresponding baseline MAD (the largest, $-$0.184 for the Transformer at 6 hr and 15 min, is just below its MAD of 0.185), and the SXR channels carry zero pooled median permutation importance in both models (Section~\ref{sec:base}), so the Slim-TSF gain cannot be attributed to the flare channels themselves; it may also reflect the altered feature sampling of the refitted forests. With SXR included in both, Slim-TSF has the higher median TSS in eight of the nine configurations (the Transformer leads only at 3 hr / 60 min), the higher precision in all nine, and the higher F1 and lower FPR in eight (FPR is tied in the ninth).

The weak SXR effect is plausible for three reasons. First, observation windows of 3--12 hr are short relative to the delays between an eruption and the arrival of $\geq$10 MeV protons: for well-connected impulsive events the parent flare may fall inside the window, but for gradual, CME-shock-dominated events the associated flare can precede the proton onset by many hours to days and lie outside it \citep{Reames2013, Cane1988}. Second, GOES SXR measurements are disk-integrated \citep{Woods2024} and carry no information on source location or magnetic connectivity, which largely determines whether accelerated protons reach the near-Earth environment \citep{Cane1988, Reames1999, DesaiGiacalone2016}, whereas in-situ electrons sample the connected population directly. Third, both models already derive most of their skill from the lowest-energy EPAM \texttt{LEFS60} bands (Section~\ref{sec:base}), which arrive $\sim$44--47 min before 10 MeV protons (Table~\ref{tab:timing}), so slower and less directly connected flare context can add little \citep{Posner2007}. We therefore regard SXR flare context as secondary to the electron flux, although a modest benefit for Slim-TSF at the 3-hr window cannot be excluded with seven folds.

\subsection{Confusion Matrix Analysis} \label{sec:cm}
\begin{figure} 
\centering 
\includegraphics[width=.99\textwidth]{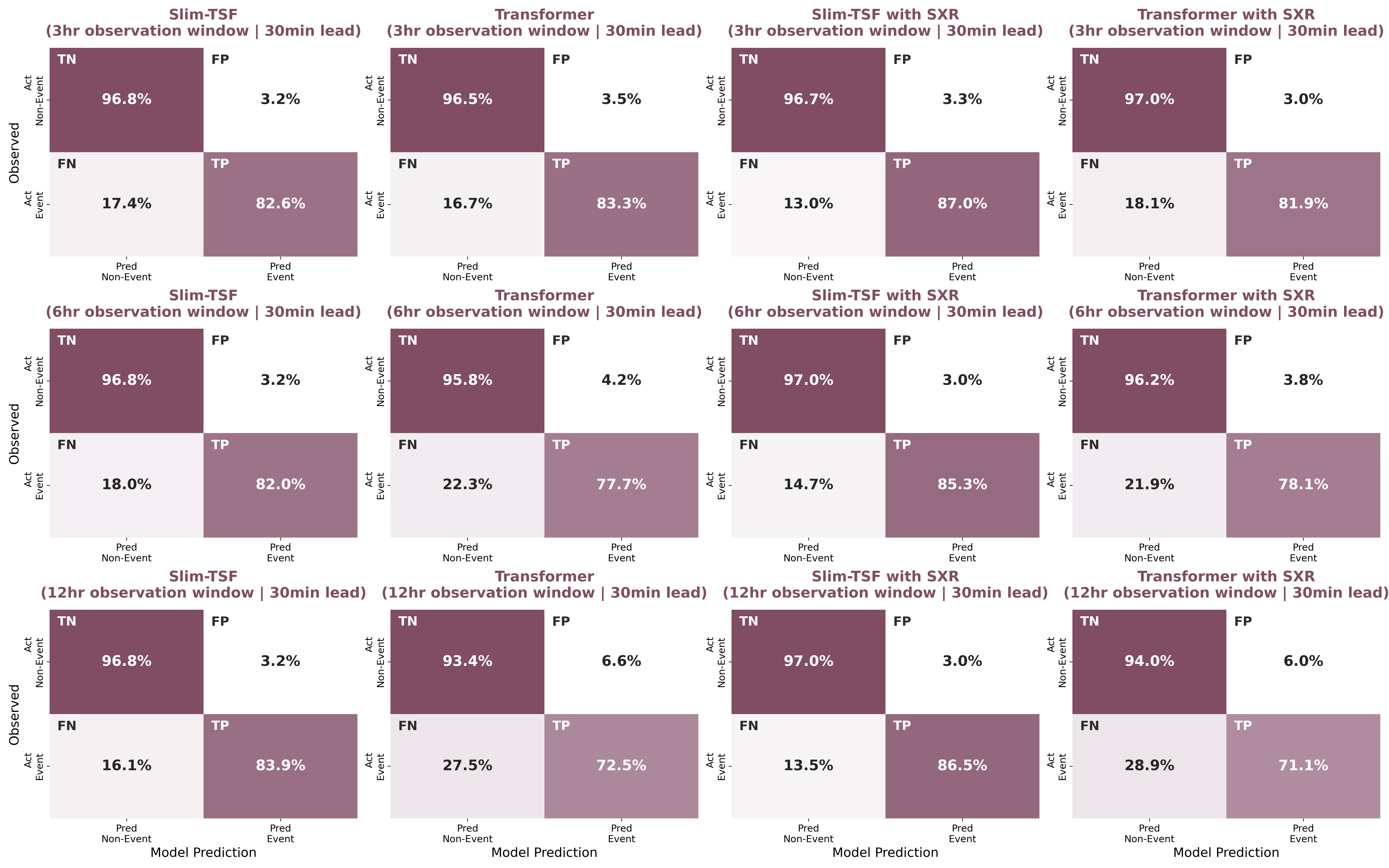} 
   \caption{Confusion matrices for the baseline (13-feature) configuration at the 30-min lead time; cell values are percentages of each observed class, with TP/TN/FP/FN counts pooled across the 30 runs and seven folds before the rates are computed, at the validation-tuned decision threshold. The first two columns show the 13-feature baseline models; the last two show the 15-feature variants with GOES SXR included.}
 \label{fig:cm}
\end{figure}

Figure~\ref{fig:cm} shows how the models categorize windows at the 30-min lead time. At 3 hr the two models recover a similar share of the positive windows (true positives; 83.3\% for the Transformer, 82.6\% for Slim-TSF), while at 6 and 12 hr Slim-TSF recovers 82.0\% and 83.9\% against 77.7\% and 72.5\%, an advantage of 4.3 and 11.4 percentage points. Slim-TSF's FP rate is 3.2\% of the negative windows at all three windows; the Transformer's is 3.5\%, 4.2\% and 6.6\%. Counted per distinct SPE (an event counts as caught if any of its positive windows raises an alarm, and is counted once for every fold in which it is tested), Slim-TSF catches 89.5\%, 87.3\% and 88.6\% of the events at 3, 6 and 12 hr, against 87.1\%, 82.6\% and 77.0\% for the Transformer, so at the event level Slim-TSF is ahead at every window, by 2.4 to 11.6 percentage points. Adding SXR shifts the Slim-TSF TP rate by +4.4, +3.3 and +2.6 percentage points and the Transformer's by $-$1.4, +0.4 and $-$1.4, with FP rates changing by at most 0.6 percentage points. The confusion matrices pool TP/TN/FP/FN counts across folds and runs before computing rates, whereas Tables~\ref{tab:skill_part1}--\ref{tab:skill_part2} report the median of the per-fold rates, so the two are not expected to coincide. Because Slim-TSF leads on recall at 6 and 12 hr and catches more events at every window, it is the stronger candidate for ``all-clear''-like predictions, where missing any single event might lead to significant consequences \citep{Sadykov2021allclear,OKeefe2024randomhivemind}. At the 3-hr window the two models are comparable, and the absence of complete fold failures makes the Transformer a reasonable choice when only a short observation window is available. One caveat applies: Slim-TSF's forests use \texttt{scikit-learn} balanced weighting (the inverse base rate, a median of 10:1 to 123:1 across configurations, Table~\ref{tab:fold_detail}), whereas the Transformer is trained at a fixed 1:5 ratio. Because the decision threshold is tuned on the validation slice in five of the seven folds, much of this difference is absorbed into the operating point, and replacing the Transformer's weights by per-fold balanced ones changes its TSS by at most 0.029 (Section~\ref{sec:balance}).

\subsection{Feature Ranking} \label{sec:base}

\begin{figure*}
\centering
\includegraphics[width=\textwidth]{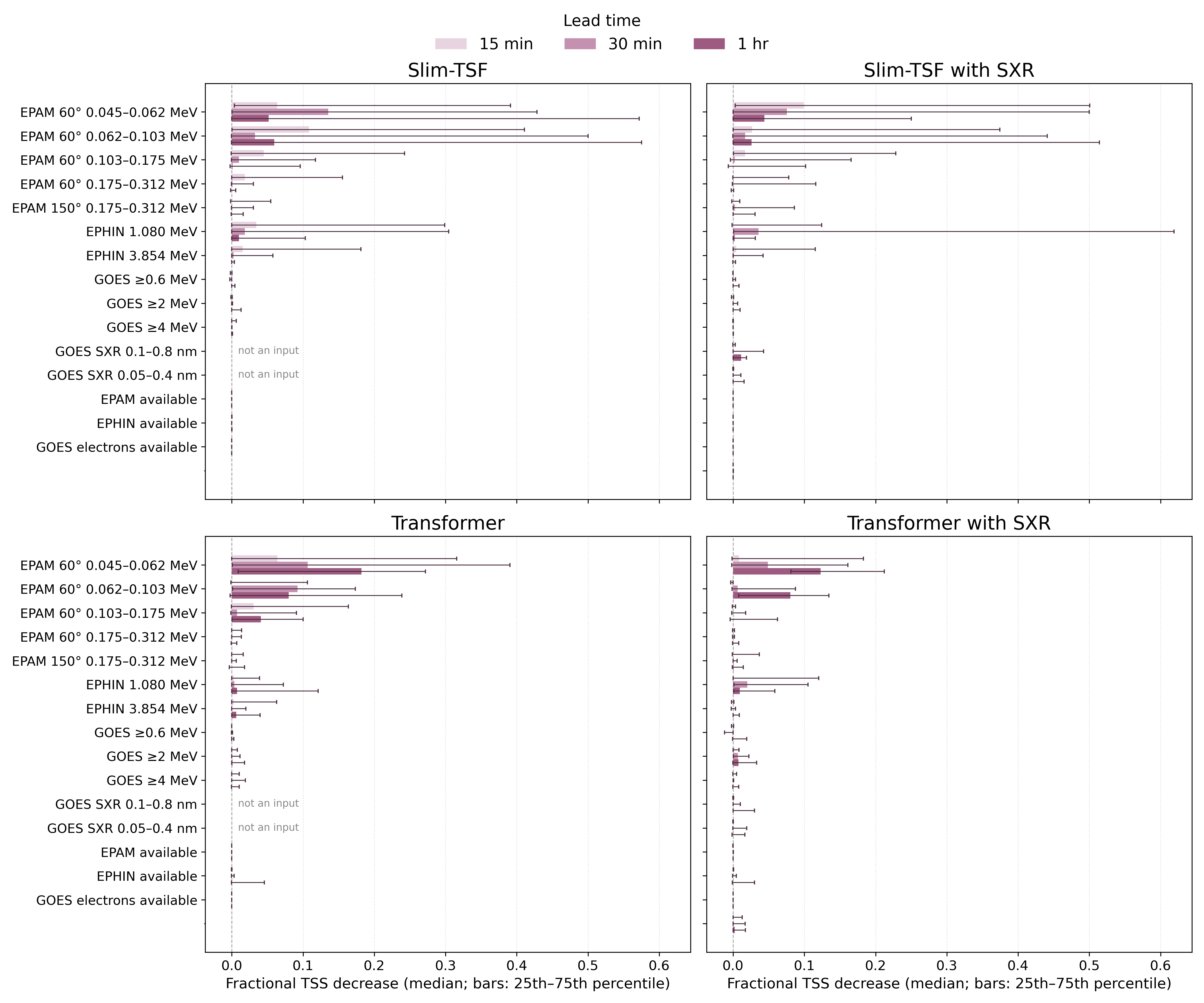}
   \caption{Median permutation importance of each individual input channel at each lead time (15, 30 and 60 min), for Slim-TSF and the Transformer, without (left column, top and bottom rows) and with GOES SXR data (right column; 15 features). Scores are the fractional TSS decreases $\mathrm{PI}_{k}/s_{k}$ of Eqs.~\ref{eq:pi-group}~and~\ref{eq:pi-final}, taken as the median over the three observation windows and the seven walk-forward folds of each. Bars show the median and whiskers the 25th to 75th percentile of these values. Scores are not constrained to sum to unity, and a value of zero means that shuffling the channel left the TSS unchanged in the typical fold.}
 \label{fi_all}
\end{figure*}
Figure~\ref{fi_all} shows the permutation importance of every input channel for the four variants on a common scale.
\begin{itemize}
    \item \textbf{Slim-TSF (baseline)} draws most of its skill from the two lowest-energy EPAM 60$^{\circ}$ bands, which are nearly tied: 0.045--0.062 MeV at $\sim$0.064 (0.064, 0.136 and 0.052 at the 15-, 30- and 60-min leads) and 0.062--0.103 MeV at $\sim$0.060 (0.109, 0.033 and 0.060). The EPHIN 1.080 MeV channel (0.018) and the 0.103--0.175 MeV band (0.010) follow, and every other channel has a pooled median of at most 0.001. The four \texttt{LEFS60} bands carry 87\% of the summed positive scores, but all single-channel scores sum to only $\sim$0.154, far below unity: any one channel can be decoupled from the label at little cost in the typical fold, because the other channels carry similar information.
    \item \textbf{Slim-TSF with SXR} keeps the same hierarchy: 0.045--0.062 MeV at $\sim$0.082, then 0.062--0.103 MeV at $\sim$0.019 and EPHIN 1.080 MeV at 0.011, with the \texttt{LEFS60} bands carrying 90\% of the summed positive scores. Both SXR channels have a pooled median of zero (at most 0.011 at any single lead time).
    \item \textbf{Transformer (baseline)} relies most on the 0.045--0.062 MeV band ($\sim$0.131), followed by 0.062--0.103 MeV ($\sim$0.066) and 0.103--0.175 MeV ($\sim$0.026), then EPHIN 1.080 MeV (0.004) and 3.854 MeV (0.003); every other channel is below 0.001 when pooled. The importance of the leading band grows with lead time (0.064, 0.107 and 0.182 at 15, 30 and 60 min). Its single-channel scores sum to $\sim$0.230, about 1.5 times Slim-TSF's.
    \item \textbf{Transformer with SXR} is dominated by the 0.045--0.062 MeV band ($\sim$0.092), followed by 0.062--0.103 MeV ($\sim$0.006), EPHIN 1.080 MeV (0.004) and GOES $\geq$2 MeV (0.003); the single-channel total is $\sim$0.106. Both SXR channels have a pooled median of zero (below 0.0005 at any lead time).
\end{itemize}
The fold distributions behind these medians are strongly skewed. For the leading 0.045--0.062 MeV band, 25--32\% of the folds show no measurable loss ($|\mathrm{PI}_k/s_k| \leq 0.005$), while the upper quartile reaches 0.20--0.50 depending on the variant. The importance of a channel is therefore concentrated in a subset of folds, and in the remaining folds the other channels compensate for its loss, consistent with the redundancy noted above. The same bands lead at all lead times: the 0.045--0.062 MeV band ranks first in ten of the twelve combinations of model variant and lead time (the 0.062--0.103 MeV band leads for Slim-TSF at 15 and 60 min), despite its slowest kinematic arrival (43.6 min lead on 10 MeV protons, Table~\ref{tab:timing}). This apparent disparity likely reflects a signal-to-noise advantage: low-energy bands have better counting statistics at onset, which lets the models resolve flux rises earlier and more reliably against the quiet-time background, so both architectures seem to weigh the shape and rate of the electron rise more heavily than first-arrival timing. The remaining \texttt{LEFS150} 0.175--0.312 MeV channel and the availability flags have a pooled median importance of zero in all four variants (at most 0.003 at any lead time), and the GOES electron channels rank below every channel with appreciable weight, consistent with the largely magnetospheric origin of the electron population GOES measures at these energies. Only single-channel permutations were evaluated, so the redundancy between channels has not been quantified with grouped permutations.

\subsection{Comparison with Established Benchmarks} \label{sec:limits}
The empirical REleASE scheme \citep{Posner2007} and the ML framework of \citet{Torres2025} provide the closest benchmarks for electron-driven forecasting, although all three approaches use different input sets and intervals. Figure~\ref{leads} marks their published F1 at the 30-min and 1-hr lead times, the only metric available from both works. Pooled over observation windows, folds and runs, all four of our model variants exceed the F1 of \citet{Torres2025} at both lead times (0.451--0.667 against 0.30 at 30 min) and that of \citet{Posner2007} at 1 hr (0.570--0.674 against 0.31). At 30 min both Slim-TSF variants also exceed the 0.56 of \citet{Posner2007} (0.588, and 0.667 with SXR), while both Transformer variants (0.451 and 0.462) fall below it. Because the studies use different intervals and positive-window definitions, this establishes that our models are in the same performance regime as established methods rather than that they are quantitatively superior. Similar to our results, \citet{Torres2025} found that adding normalized GOES X-ray features to their network yielded no significant gain over purely electron-driven configurations.

Both benchmarks lose skill from 30 min to 1 hr (F1 from 0.30 to 0.16 for \citealt{Torres2025} and from 0.56 to 0.31 for \citealt{Posner2007}), whereas our F1 rises (e.g., 0.588 to 0.651 for Slim-TSF and 0.451 to 0.570 for the Transformer) and our TSS changes little. We attribute this mainly to the forecast target. REleASE and the networks of \citet{Torres2025} predict the proton flux itself at a time 30 or 60 min ahead, and a point forecast becomes harder the further ahead it is made. Our target is instead whether the 10 pfu crossing occurs anywhere within the lead time, so a longer lead time adds issue times on which the crossing is still ahead, rather than pushing every forecast further into the future. A longer lead time also raises the positive fraction of the windows (from 0.7\% at 3 hr / 15 min to 8.9\% at 12 hr / 60 min, Table~\ref{tab:samples}), which raises precision and therefore F1 independently of skill, whereas TSS, which does not depend on the base rate, stays nearly flat. The differing lead-time trends therefore reflect the different target definitions and the base-rate dependence of F1 rather than a physical difference between the methods.


\section{Summary and Conclusions}\label{sec:discussion}
In-situ sub-relativistic electron precursors enable useful short-term SPE forecasts across 3--12 hr observation windows and 15--60 min lead times. The key findings are:
\begin{itemize}
    \item \textbf{Slim-TSF is ahead at 6 and 12 hr and comparable at 3 hr.} Its median TSS is 0.702--0.837 at 3 hr, 0.773--0.791 at 6 hr and 0.789--0.897 at 12 hr, at recall 0.727--1.000, precision 0.178--0.607 and FPR $\leq$0.023 in eight of nine configurations (0.061 at 12 hr / 60 min). Its 12-hr advantage follows the more balanced 12-hr training set rather than the longer window (Section~\ref{sec:balance}), and its main weakness is three 3-hr folds in which it fails outright. This makes it the natural choice for all-clear-like forecasting, where missing an event costs more than an alert that needs downstream verification.
    \item \textbf{The Transformer is competitive at 3 hr and declines with window length.} Its median TSS is 0.797--0.873 at 3 hr, 0.624--0.684 at 6 hr and 0.586--0.641 at 12 hr, at recall 0.675--0.915 and precision 0.185--0.585. Its FPR is the higher of the two in eight of nine configurations, and it is less reproducible from run to run (median seed-to-seed standard deviation of the fold TSS 0.08 against 0.01). The decline comes mainly from averaging over the window: with last-step pooling its 12-hr TSS rises from 0.59 to 0.87, whereas per-fold class weights change it by at most 0.029 and four more training epochs by at most 0.03 at 12 hr and 0.08 at 6 hr (Section~\ref{sec:balance}, Appendix~\ref{app:convergence}).
    \item \textbf{Lead time has little effect.} Pooled median TSS varies by 0.046 (Slim-TSF) and 0.082 (Transformer) across the 15-, 30- and 60-min leads, against a fold-and-run MAD of 0.17--0.22 within each lead time.
    \item \textbf{SXR has at most a small effect.} It produces no consistent change in the Transformer (median TSS shifts of $-$0.184 to +0.058 with mixed sign) and a small Slim-TSF gain concentrated at 3 hr with 15- and 30-min leads (mean paired fold change +0.030). All shifts lie within the baseline MAD, and the SXR channels have zero pooled median permutation importance in both models.
    \item \textbf{Sub-relativistic electrons dominate the feature hierarchies.} Both models draw most of their skill from the lowest-energy EPAM \texttt{LEFS60} bands; the 0.045--0.062 MeV band is the strongest single channel in ten of twelve combinations of model variant and lead time, and the \texttt{LEFS60} bands account for 87--97\% of the summed positive scores. The EPHIN 1.080 MeV channel follows (pooled $\leq$0.018); the remaining \texttt{LEFS150} channel, the GOES electron channels, SXR and the availability flags are negligible.
    \item \textbf{Both models show operational value.} Pooled over windows, the median F1 of all four variants exceeds both published benchmarks at the 1-hr lead time (0.570--0.674) and that of \citet{Torres2025} at 30 min (0.451--0.667), and at 30 min both Slim-TSF variants also exceed that of \citet{Posner2007}, subject to the caveats of Section~\ref{sec:limits}. Model choice is dictated by the observation window available and by alert-tolerance requirements: Slim-TSF produces fewer missed events at 6 and 12 hr, whereas the Transformer is a reasonable choice for short, 3-hr observation windows, where the two are comparable.
\end{itemize}

\subsection{Operational outlook}
The kinematic margins of Table~\ref{tab:timing} are set by particle transport rather than by either model, so any measurement that samples a faster or earlier-arriving population can potentially extend the warning time. New upstream capabilities, like the recently deployed Space Weather Follow On at L1 (SWFO-L1) observatory and its Suprathermal Ion Sensor (STIS), provide real-time, high-resolution measurements at the L1 point. Incorporating suprathermal fluxes as an additional precursor, again leveraging the lead-time advantage these particles hold over hazardous SEP populations, could move both the Slim-TSF and Transformer frameworks toward a more comprehensive, physics-informed forecasting system. Neither architecture is tied to a particular vantage point or instrument, so both can potentially be extended to operational pipelines drawing on upstream observatories as those measurements become available.

\appendix

\section{Fold Composition}\label{app:samples}
\begin{deluxetable*}{ccrrrrrr}
\tablecaption{Window-level sample distribution with onset-only labeling across observation windows and lead times. A window is labeled positive only if a new event begins inside the lead-time interval that follows it \textit{and} no event was already active during the observation period. Positive windows are taken at every 5-min issue time, so one onset yields up to 3, 6 or 12 overlapping positive windows for the 15-, 30- and 60-min leads; negative windows are non-overlapping and are quiet through the lead time and a 3-hr buffer beyond it. The last column is the number of the 212 onsets in the record that supply at least one positive window.\label{tab:samples}}
\tablehead{\colhead{Obs. Period} & \colhead{Lead time} & \colhead{Windows} & \colhead{Positive} & \colhead{Negative} & \colhead{Positive} & \colhead{Imbalance} & \colhead{SPEs} \\
\colhead{(hr)} & \colhead{(min)} & \colhead{(N)} & \colhead{(N)} & \colhead{(N)} & \colhead{(\%)} & \colhead{(neg:pos)} & \colhead{(N)}}
\startdata
\multirow{3}{*}{3} & 15 & 57,324 & 400 & 56,924 & 0.698 & 142:1 & 134 \\
& 30 & 57,507 & 789 & 56,718 & 1.372 & 72:1 & 134 \\
& 60 & 57,954 & 1,550 & 56,404 & 2.675 & 36:1 & 134 \\
\cmidrule(lr){1-8}
\multirow{3}{*}{6} & 15 & 27,772 & 348 & 27,424 & 1.253 & 79:1 & 116 \\
& 30 & 28,012 & 693 & 27,319 & 2.474 & 39:1 & 116 \\
& 60 & 28,532 & 1,372 & 27,160 & 4.809 & 20:1 & 116 \\
\cmidrule(lr){1-8}
\multirow{3}{*}{12} & 15 & 13,043 & 315 & 12,728 & 2.415 & 40:1 & 105 \\
& 30 & 13,304 & 627 & 12,677 & 4.713 & 20:1 & 105 \\
& 60 & 13,840 & 1,238 & 12,602 & 8.945 & 10:1 & 105 \\
\enddata
\end{deluxetable*}

\begin{deluxetable*}{ccrrrrrr}
\tablecaption{Composition of the seven walk-forward folds of each configuration (identical for both models, for every run and with or without SXR). Entries are medians over the folds, with the minimum and maximum across folds in brackets. The training imbalance is the ratio of negative to positive windows in the training partition (fit and validation slices together). The last column counts the folds whose test partition contains only two distinct events.\label{tab:fold_detail}}
\tablehead{\colhead{Obs. Period} & \colhead{Lead time} & \colhead{Train} & \colhead{Test} & \colhead{Test positive} & \colhead{Unique Test} & \colhead{Train imbalance} & \colhead{Folds with} \\
\colhead{(hr)} & \colhead{(min)} & \colhead{(N)} & \colhead{(N)} & \colhead{windows (N)} & \colhead{events (N)} & \colhead{(neg:pos)} & \colhead{2 test events}}
\startdata
\multirow{3}{*}{3} & 15 & 10,030 & 4,299 & 30 [6--71] & 10 [2--24] & 123:1 [62:1--1671:1] & 3 of 7 \\
& 30 & 10,062 & 4,312 & 60 [12--140] & 10 [2--24] & 61:1 [31:1--838:1] & 3 of 7 \\
& 60 & 10,140 & 4,346 & 114 [24--275] & 10 [2--24] & 31:1 [15:1--422:1] & 3 of 7 \\
\cmidrule(lr){1-8}
\multirow{3}{*}{6} & 15 & 4,859 & 2,082 & 21 [6--66] & 7 [2--22] & 76:1 [33:1--809:1] & 3 of 7 \\
& 30 & 4,901 & 2,100 & 42 [12--132] & 7 [2--22] & 38:1 [16:1--408:1] & 3 of 7 \\
& 60 & 4,992 & 2,139 & 72 [20--271] & 6 [2--23] & 19:1 [8:1--207:1] & 3 of 7 \\
\cmidrule(lr){1-8}
\multirow{3}{*}{12} & 15 & 2,281 & 977 & 21 [6--57] & 7 [2--19] & 38:1 [17:1--379:1] & 3 of 7 \\
& 30 & 2,327 & 997 & 42 [12--120] & 7 [2--20] & 19:1 [8:1--193:1] & 3 of 7 \\
& 60 & 2,421 & 1,037 & 72 [24--231] & 6 [2--20] & 10:1 [4:1--100:1] & 3 of 7 \\
\enddata
\end{deluxetable*}
\begin{deluxetable}{lrrrrrrr}
\tablecaption{Percentage of test-partition time steps still missing after the 1-hr forward fill, per walk-forward fold (6-hr window, 30-min lead; the other configurations are similar). These steps are set to the training mean and flagged as unavailable. The EPHIN series used here ends on 4 October 2017, so its channels are flagged for almost the whole test partition of the last fold.\label{tab:fold_coverage}}
\tablehead{\colhead{Channel} & \colhead{F1} & \colhead{F2} & \colhead{F3} & \colhead{F4} & \colhead{F5} & \colhead{F6} & \colhead{F7}}
\startdata
EPAM LEFS60 (all bands) & 0 & 0 & 0 & 0 & 1 & 3 & 2 \\
EPAM LEFS150 0.175--0.312 MeV & 0 & 0 & 0 & 0 & 1 & 3 & 2 \\
EPHIN 1.080 MeV & 2 & 6 & 9 & 11 & 4 & 10 & 98 \\
EPHIN 3.854 MeV & 2 & 6 & 9 & 11 & 4 & 8 & 98 \\
GOES $\geq$0.6, $\geq$2, $\geq$4 MeV & 0 & 0 & 0 & 0 & 0 & 0 & 7 \\
GOES XRS 0.1--0.8 nm & 0 & 0 & 1 & 0 & 0 & 1 & 3 \\
GOES XRS 0.05--0.4 nm & 0 & 0 & 1 & 9 & 14 & 6 & 0 \\
\enddata
\end{deluxetable}

Tables~\ref{tab:samples} and \ref{tab:fold_detail} enumerate the windows and the folds summarized in Section~\ref{meth}. The folds with the fewest test events dominate the fold-to-fold spread quoted as MAD in Tables~\ref{tab:skill_part1} and \ref{tab:skill_part2}: the test partition of three of the seven folds of every configuration (27 of 63 in total) holds only two distinct SPEs, which supply only 6 to 24 positive windows depending on the lead time, so catching or missing a single event moves the fold TSS by tens of percent. The fourth fold of every configuration, in turn, has only two events in its training partition, and in it and in the third fold the validation slice is empty, so the static 0.30 threshold is used instead of a validated one. Table~\ref{tab:fold_coverage} lists, per fold, the share of test-partition steps that remain missing after the 1-hr forward fill; the EPHIN channels are flagged as unavailable for almost the entire test partition of the last fold.

\section{Transformer Training Budget}\label{app:convergence}
\begin{figure*}[t]
\centering
\includegraphics[width=.8\textwidth]{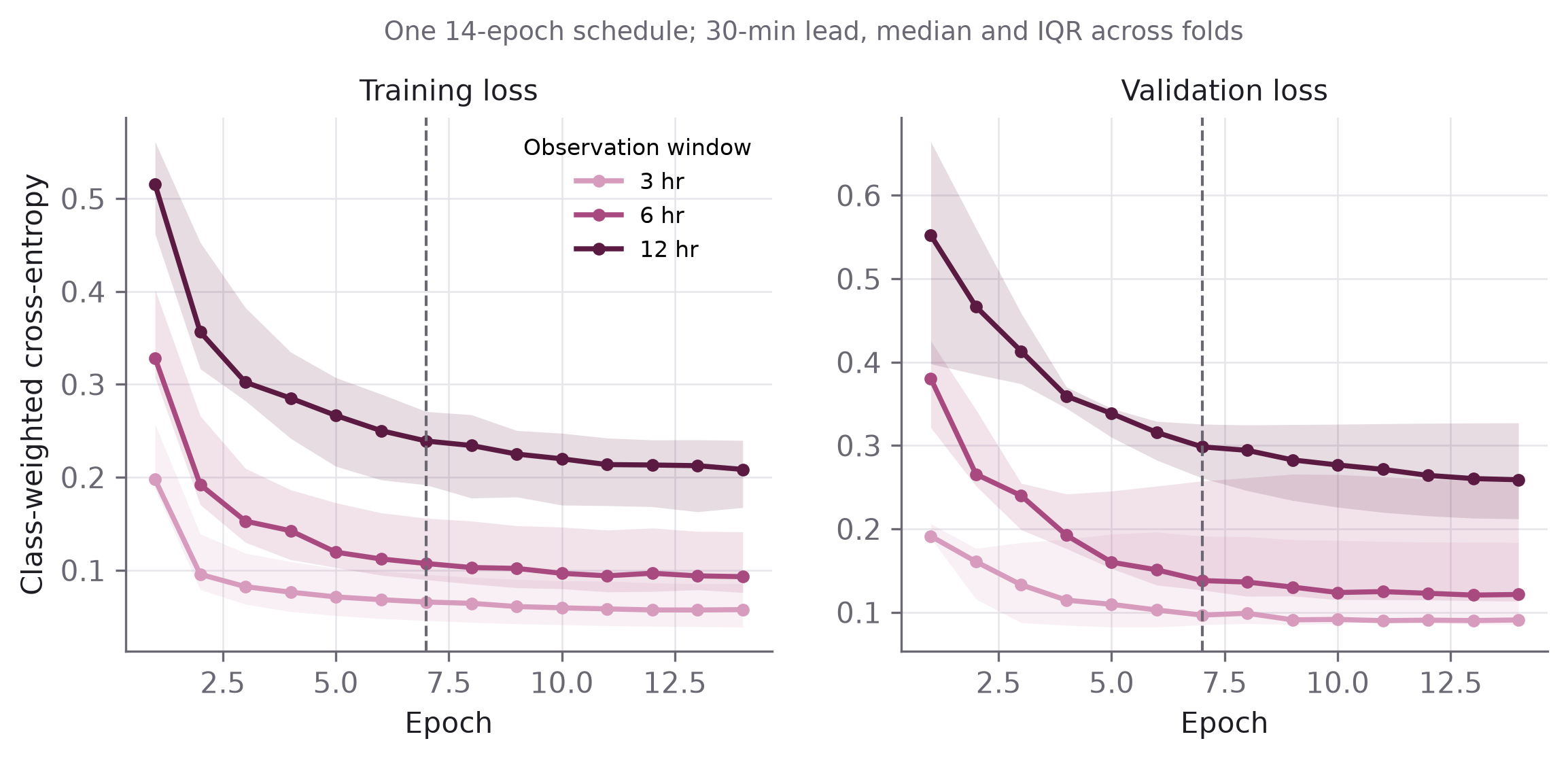}
\caption{Transformer training (left) and validation (right) class-weighted cross-entropy over a single 14-epoch schedule, for the baseline configuration at the 30-minute lead time. Lines are medians over the seven walk-forward folds of each observation window (3-, 6- and 12-hr), shaded bands give the interquartile range across folds, and the dashed line marks epoch 7. The budget adopted for the reported results is 10 epochs.\label{fig:convergence}}
\label{fig:loss}
\end{figure*}

The ten-epoch budget was fixed rather than selected per configuration. To assess its cost, we ran a diagnostic scan of epoch budgets (3, 5, 7, 10 and 14 epochs, each trained with its own complete learning-rate schedule) on the same walk-forward folds, inputs and preprocessing as the main experiments, with three seeds. Figure~\ref{fig:convergence} shows the class-weighted loss of the 14-epoch runs at the 30-min lead time: the training loss keeps decreasing slowly throughout, while the validation loss has flattened by epoch 9--10 at 3 and 6 hr (it is only 0.7\% and 1.9\% lower at epoch 14 than at epoch 10) but is still decreasing slowly at 12 hr (6.4\% lower at epoch 14). In an epoch scan relative to the 7-epoch reference, ten epochs raised the mean test TSS in every one of the nine configurations, by 0.015 to 0.110, and fourteen epochs by 0.017 to 0.130, while three and five epochs lowered it by up to 0.19 and 0.08. Relative to the adopted ten epochs, fourteen epochs change the mean test TSS by $-$0.005 to +0.022 at 3 hr, +0.012 to +0.075 at 6 hr and +0.019 to +0.030 at 12 hr; with seven folds per configuration, only the 12-hr gains at the 30- and 60-min leads are (narrowly) distinguishable from zero at the paired 95\% level. The reported Transformer scores at 6 and 12 hr may therefore understate the skill this architecture can reach by a few hundredths in TSS, and by up to 0.08 at 6 hr with a 30-min lead. This is small compared with the median-TSS gap to Slim-TSF at 12 hr (0.17--0.30) and with the gain from last-step pooling (Section~\ref{sec:balance}), but of the same order as the gap at 6 hr (0.09--0.17). We retained a fixed budget because a uniform allocation keeps training cost identical across all window--lead--fold cells and avoids time-consuming per-configuration tuning.

\section{Imbalance and Pooling Ablation}\label{app:ablation}
\begin{deluxetable*}{llcccrr}
\tablecaption{Imbalance and pooling ablation at the 30-min lead time (Appendix~\ref{app:ablation}). Median TSS\,$\pm$\,MAD and median FPR over the 30 runs and seven walk-forward folds of the main experiments, on the unchanged test partitions; the baseline rows are therefore the 30-min entries of Table~\ref{tab:skill_part1}. Training negatives and the negative-to-positive ratio are medians over the folds.\label{tab:ablation}}
\tablehead{\colhead{Model} & \colhead{Setting} & \colhead{Window (hr)} & \colhead{TSS} & \colhead{FPR} & \colhead{Train neg.} & \colhead{Neg:pos}}
\startdata
Transformer & baseline & 3 & 0.873\,$\pm$\,0.121 & 0.020 & 9836 & 61:1 \\
 & baseline & 6 & 0.677\,$\pm$\,0.167 & 0.036 & 4709 & 38:1 \\
 & baseline & 12 & 0.586\,$\pm$\,0.201 & 0.036 & 2147 & 19:1 \\
 & (a) negatives subsampled to the 12-hr ratio & 3 & 0.830\,$\pm$\,0.162 & 0.035 & 3634 & 20:1 \\
 & (b) negatives on the 3-hr stride & 12 & 0.641\,$\pm$\,0.215 & 0.022 & 7387 & 71:1 \\
 & (c) per-fold balanced class weights & 3 & 0.878\,$\pm$\,0.116 & 0.033 & 9836 & 61:1 \\
 & (c) per-fold balanced class weights & 6 & 0.681\,$\pm$\,0.158 & 0.042 & 4709 & 38:1 \\
 & (c) per-fold balanced class weights & 12 & 0.615\,$\pm$\,0.194 & 0.040 & 2147 & 19:1 \\
 & (d) last-step pooling & 3 & 0.916\,$\pm$\,0.083 & 0.011 & 9836 & 61:1 \\
 & (d) last-step pooling & 12 & 0.869\,$\pm$\,0.127 & 0.024 & 2147 & 19:1 \\
\tableline
Slim-TSF & baseline & 3 & 0.785\,$\pm$\,0.213 & 0.016 & 9836 & 61:1 \\
 & baseline & 6 & 0.776\,$\pm$\,0.216 & 0.014 & 4709 & 38:1 \\
 & baseline & 12 & 0.880\,$\pm$\,0.113 & 0.010 & 2147 & 19:1 \\
 & (a) negatives subsampled to the 12-hr ratio & 3 & 0.941\,$\pm$\,0.057 & 0.024 & 3634 & 20:1 \\
 & (b) negatives on the 3-hr stride & 12 & 0.770\,$\pm$\,0.219 & 0.012 & 7387 & 71:1 \\
\enddata
\end{deluxetable*}

To separate training-set size, class ratio and pooling (Section~\ref{sec:balance}), we reran both models at the 30-min lead with the 30 seeds of the main experiments on the unchanged walk-forward test folds, so that the baseline rows of Table~\ref{tab:ablation} are the 30-min entries of Table~\ref{tab:skill_part1}. Only the fitting part of each training partition is changed; the validation slice, threshold rule and test partitions are those of the main experiments. In (a), training negatives of the 3-hr windows are randomly subsampled to the class ratio of the corresponding 12-hr fold (median 20:1 instead of 61:1). In (b), negatives of the 12-hr windows are drawn at the 3-hr stride under the same quiet-time and buffer rules (71:1 instead of 19:1); these negatives overlap, so the effective sample is smaller than the count. In (c), the Transformer's fixed 1:5 class weights are replaced by per-fold inverse-frequency weights. In (d), the Transformer classifies from the encoder state at the last time step instead of the time average. Slim-TSF repeats its hyperparameter search for (a) and (b). Table~\ref{tab:ablation} lists the results.

\clearpage
\section*{Acknowledgements}
This work used Anthropic's Claude in a limited capacity; specifically to help structure the hyperparameter grid for Slim-TSF configurations and in interpreting Transformer convergence diagnostics. Both cases emphasized on keeping the parameter search and model training costs tractable across the full set of window-lead time configurations. All model configurations, figures, and numerical results were produced and verified by the author.

\bibliography{sample701}{}

@article{Laurenzaetal2009,
  author  = {Laurenza, M. and Cliver, E. W. and Hewitt, J. and Storini, M. and Ling, A. G. and Balch, C. C. and Kaiser, M. L.},
  title   = {A technique for short‐term warning of solar energetic particle events based on flare location, flare size, and evidence of particle escape},
  journal = {Space Weather},
  year    = {2009},
  volume  = {7},
  doi     = {10.1029/2007sw000379}
}

@article{Huangetal2012,
  author  = {Huang, Xin and Wang, Hua-Ning and Li, Le-Ping},
  title   = {Ensemble prediction model of solar proton events associated with solar flares and coronal mass ejections},
  journal = {Research in Astronomy and Astrophysics},
  year    = {2012},
  volume  = {12},
  number  = {3},
  pages   = {313--321},
  doi     = {10.1088/1674-4527/12/3/007}
}

@article{Swalwelletal2017,
  author  = {Swalwell, B. and Dalla, S. and Walsh, R. W.},
  title   = {Solar Energetic Particle Forecasting Algorithms and Associated False Alarms},
  journal = {Solar Physics},
  year    = {2017},
  volume  = {292},
  doi     = {10.1007/s11207-017-1196-y}
}

@article{Torres2025,
  author = {Torres, Jesse and Chan, Philip K. and Zhao, Lulu and Zhang, Ming},
  title = {A Machine Learning Approach to Predicting SEP Proton Intensity and Events Using Time Series of Relativistic Electron Measurements},
  journal = {Space Weather},
  volume = {23},
  number = {e2024SW003921},
  year = {2025},
  doi = {10.1029/2024SW003921}
}

@article{Posner2007,
  author = {Posner, Arik},
  title = {Up to 1-hour forecasting of radiation hazards from solar energetic ion events with relativistic electrons},
  journal = {Space Weather},
  volume = {5},
  number = {S05001},
  year = {2007},
  doi = {10.1029/2006SW000268}
}

@article{Breiman2001,
  author  = {Breiman, Leo},
  title   = {Random Forests},
  journal = {Machine Learning},
  year    = {2001},
  volume  = {45},
  number  = {1},
  pages   = {5--32},
  doi     = {10.1023/A:1010933404324}
}

@article{memsep,
author = {Moreland, Kimberly and Dayeh, Maher A. and Bain, Hazel M. and Chatterjee, Subhamoy and Muñoz-Jaramillo, Andrés and Hart, Samuel T.},
title = {MEMPSEP-III. A Machine Learning-Oriented Multivariate Data Set for Forecasting the Occurrence and Properties of Solar Energetic Particle Events Using a Multivariate Ensemble Approach},
journal = {Space Weather},
volume = {22},
number = {9},
pages = {e2023SW003765},
doi = {https://doi.org/10.1029/2023SW003765},
url = {https://agupubs.onlinelibrary.wiley.com/doi/abs/10.1029/2023SW003765},
eprint = {https://agupubs.onlinelibrary.wiley.com/doi/pdf/10.1029/2023SW003765},
note = {e2023SW003765 2023SW003765},
year = {2024}
}

@article{Vaswani2017,
  author  = {Vaswani, Ashish and Shazeer, Noam and Parmar, Niki and Uszkoreit, Jakob and Jones, Llion and Gomez, Aidan N and Kaiser, Lukasz and Polosukhin, Illia},
  title   = {Attention is all you need},
  journal = {Advances in Neural Information Processing Systems},
  volume  = {30},
  year    = {2017},
  url     = {https://arxiv.org/abs/1706.03762}
}

@inproceedings{SLIMTSF2024,
  author    = {Ji, Anli and Pandey, Chetraj and Aydin, Berkay},
  title     = {Towards Hybrid Embedded Feature Selection and Classification Approach with Slim-TSF},
  booktitle = {Proceedings of the 26th International Conference on Big Data Analytics and Knowledge Discovery (DaWaK 2024)},
  year      = {2024},
  series    = {Lecture Notes in Computer Science},
  publisher = {Springer},
  volume    = {14913},
  doi       = {10.1007/978-3-031-68323-7_7}
}

@article{kahler2001,
  author  = {Kahler, Stephen W.},
  title   = {The Correlation between Solar Energetic Particle and Coronal Mass Ejection Speeds},
  journal = {Journal of Geophysical Research},
  volume  = {106},
  pages   = {20947--20956},
  year    = {2001},
  doi     = {10.1029/2000JA000345}
}

@article{lin1974,
  author  = {Lin, Robert P.},
  title   = {Nonrelativistic Solar Electrons},
  journal = {Space Science Reviews},
  volume  = {16},
  pages   = {189--223},
  year    = {1974},
  doi     = {10.1007/BF00185766}
}

@ARTICLE{Bobra2014SHARP,
       author = {{Bobra}, M.~G. and {Sun}, X. and {Hoeksema}, J.~T. and {Turmon}, M. and {Liu}, Y. and {Hayashi}, K. and {Barnes}, G. and {Leka}, K.~D.},
        title = "{The Helioseismic and Magnetic Imager (HMI) Vector Magnetic Field Pipeline: SHARPs - Space-Weather HMI Active Region Patches}",
      journal = {Solar Physics},
         year = 2014,
        month = sep,
       volume = {289},
       number = {9},
        pages = {3549-3578},
          doi = {10.1007/s11207-014-0529-3},
archivePrefix = {arXiv},
       eprint = {1404.1879},
 primaryClass = {astro-ph.SR},
       adsurl = {https://ui.adsabs.harvard.edu/abs/2014SoPh..289.3549B}
}

@ARTICLE{Sadykov2021allclear,
       author = {{Sadykov}, Viacheslav and {Kosovichev}, Alexander and {Kitiashvili}, Irina and {Oria}, Vincent and {Nita}, Gelu M and {Illarionov}, Egor and {O'Keefe}, Patrick and {Jiang}, Yucheng and {Fereira}, Sheldon and {Ali}, Aatiya},
        title = "{Prediction of Solar Proton Events with Machine Learning: Comparison with Operational Forecasts and ``All-Clear'' Perspectives}",
      journal = {arXiv e-prints},
         year = 2021,
        month = jul,
          eid = {arXiv:2107.03911},
        pages = {arXiv:2107.03911},
          doi = {10.48550/arXiv.2107.03911},
archivePrefix = {arXiv},
       eprint = {2107.03911},
 primaryClass = {astro-ph.SR},
       adsurl = {https://ui.adsabs.harvard.edu/abs/2021arXiv210703911S}
}

@ARTICLE{OKeefe2024randomhivemind,
       author = {{O'Keefe}, Patrick M. and {Sadykov}, Viacheslav and {Kosovichev}, Alexander and {Kitiashvili}, Irina N. and {Oria}, Vincent and {Nita}, Gelu M. and {Francis}, Fraila and {Chong}, Chun-Jie and {Kosovich}, Paul and {Ali}, Aatiya and {Marroquin}, Russell D.},
        title = "{The Random Hivemind: An ensemble deep learning application to the solar energetic particle prediction problem}",
      journal = {Advances in Space Research},
         year = 2024,
        month = dec,
       volume = {74},
       number = {12},
        pages = {6252-6263},
          doi = {10.1016/j.asr.2024.04.044},
archivePrefix = {arXiv},
       eprint = {2303.08092},
 primaryClass = {astro-ph.SR},
       adsurl = {https://ui.adsabs.harvard.edu/abs/2024AdSpR..74.6252O}
}

@misc{zenodo,
  author       = {Dr{\"o}ge, Henrik and Heber, Bernd},
  title        = {{SOHO EPHIN corrected electron dataset}},
  month        = jul,
  year         = 2024,
  publisher    = {Zenodo},
  version      = {v1.0},
  doi          = {10.5281/zenodo.17286228},
  url          = {https://doi.org/10.5281/zenodo.17286228}
}

@techreport{RodriguezATBD2014,
  author      = {Rodriguez, J. V.},
  title       = {{GOES} 13--15 {EPEAD} Science-Quality Electron Fluxes:
                 Algorithm Theoretical Basis Document, Version 1.0},
  institution = {NOAA National Geophysical Data Center},
  year        = {2014},
  url         = {https://www.ngdc.noaa.gov/stp/satellite/goes/doc/EPEAD_Electron_Science_Reprocessing_ATBD_v1.0.pdf}
}

@article{pedregosa,
  author  = {Pedregosa, F. and Varoquaux, G. and Gramfort, A. and Michel, V. and
             Thirion, B. and Grisel, O. and Blondel, M. and Prettenhofer, P. and
             Weiss, R. and Dubourg, V. and Vanderplas, J. and Passos, A. and
             Cournapeau, D. and Brucher, M. and Perrot, M. and Duchesnay, E.},
  title   = {Scikit-learn: Machine Learning in {P}ython},
  journal = {Journal of Machine Learning Research},
  volume  = {12},
  pages   = {2825--2830},
  year    = {2011}
}

@inproceedings{jamieson,
  author        = {Jamieson, Kevin and Talwalkar, Ameet},
  title         = {Non-stochastic Best Arm Identification and Hyperparameter Optimization},
  booktitle     = {Proceedings of the 19th International Conference on Artificial
                   Intelligence and Statistics (AISTATS)},
  series        = {Proceedings of Machine Learning Research},
  volume        = {51},
  pages         = {240--248},
  year          = {2016},
  eprint        = {1502.07943},
  archivePrefix = {arXiv},
  doi           = {10.48550/arXiv.1502.07943}
}

@article{fisher2019,
  author        = {Fisher, Aaron and Rudin, Cynthia and Dominici, Francesca},
  title         = {All Models are Wrong, but Many are Useful: Learning a Variable's
                   Importance by Studying an Entire Class of Prediction Models
                   Simultaneously},
  journal       = {Journal of Machine Learning Research},
  volume        = {20},
  number        = {177},
  pages         = {1--81},
  year          = {2019},
  eprint        = {1801.01489},
  archivePrefix = {arXiv},
  doi           = {10.48550/arXiv.1801.01489}
}

@inproceedings{loshchilov2019,
  author        = {Loshchilov, Ilya and Hutter, Frank},
  title         = {Decoupled Weight Decay Regularization},
  booktitle     = {International Conference on Learning Representations (ICLR)},
  year          = {2019},
  eprint        = {1711.05101},
  archivePrefix = {arXiv},
  doi           = {10.48550/arXiv.1711.05101}
}

@inproceedings{wolf2020,
  author    = {Wolf, Thomas and Debut, Lysandre and Sanh, Victor and Chaumond, Julien
               and Delangue, Clement and Moi, Anthony and Cistac, Pierric and
               Rault, Tim and Louf, R{\'e}mi and Funtowicz, Morgan and Davison, Joe
               and Shleifer, Sam and von Platen, Patrick and Ma, Clara and
               Jernite, Yacine and Plu, Julien and Xu, Canwen and {Le Scao}, Teven
               and Gugger, Sylvain and Drame, Mariama and Lhoest, Quentin and
               Rush, Alexander M.},
  title     = {Transformers: State-of-the-Art Natural Language Processing},
  booktitle = {Proceedings of the 2020 Conference on Empirical Methods in Natural
               Language Processing: System Demonstrations},
  pages     = {38--45},
  year      = {2020},
  publisher = {Association for Computational Linguistics},
  doi       = {10.18653/v1/2020.emnlp-demos.6}
}

@article{Reames1999,
  author  = {Reames, Donald V.},
  title   = {Particle Acceleration at the Sun and in the Heliosphere},
  journal = {Space Science Reviews},
  volume  = {90},
  pages   = {413--491},
  year    = {1999},
  doi     = {10.1023/A:1005105831781}
}

@article{Reames2013,
  author  = {Reames, Donald V.},
  title   = {The Two Sources of Solar Energetic Particles},
  journal = {Space Science Reviews},
  volume  = {175},
  pages   = {53--92},
  year    = {2013},
  doi     = {10.1007/s11214-013-9958-9}
}

@article{Cane1988,
  author  = {Cane, Hilary V. and Reames, Donald V. and von Rosenvinge, Tycho T.},
  title   = {The Role of Interplanetary Shocks in the Longitude Distribution of Solar Energetic Particles},
  journal = {Journal of Geophysical Research: Space Physics},
  volume  = {93},
  number  = {A9},
  pages   = {9555--9567},
  year    = {1988},
  doi     = {10.1029/JA093iA09p09555}
}

@article{DesaiGiacalone2016,
  author  = {Desai, Mihir and Giacalone, Joe},
  title   = {Large Gradual Solar Energetic Particle Events},
  journal = {Living Reviews in Solar Physics},
  volume  = {13},
  pages   = {3},
  year    = {2016},
  doi     = {10.1007/s41116-016-0002-5}
}

@article{Woods2024,
  author  = {Woods, Thomas N. and Eden, Thomas and Eparvier, Francis G. and Jones, Andrew R.
             and Woodraska, Donald L. and Chamberlin, Phillip C. and Machol, Janet L.},
  title   = {{GOES-R} Series X-Ray Sensor ({XRS}): 1. Design and Pre-Flight Calibration},
  journal = {Journal of Geophysical Research: Space Physics},
  volume  = {129},
  number  = {6},
  pages   = {e2024JA032925},
  year    = {2024},
  doi     = {10.1029/2024JA032925}
}

@ARTICLE{parker,
       author = {{Parker}, E. N.},
        title = "{Dynamics of the Interplanetary Gas and Magnetic Fields}",
      journal = {\apj},
         year = 1958,
        month = nov,
       volume = {128},
        pages = {664},
          doi = {10.1086/146579},
       adsurl = {https://ui.adsabs.harvard.edu/abs/1958ApJ...128..664P}
}

@ARTICLE{me,
       author = {{Ali}, Aatiya and {Sadykov}, Viacheslav and {Kosovichev}, Alexander and {Kitiashvili}, Irina N. and {Oria}, Vincent and {Nita}, Gelu M. and {Illarionov}, Egor and {O'Keefe}, Patrick M. and {Francis}, Fraila and {Chong}, Chun-Jie and {Kosovich}, Paul and {Marroquin}, Russell D.},
        title = "{Predicting Solar Proton Events of Solar Cycles 22-24 Using GOES Proton and Soft-X-Ray Flux Features}",
      journal = {\apjs},
         year = 2024,
        month = jan,
       volume = {270},
       number = {1},
          eid = {15},
        pages = {15},
          doi = {10.3847/1538-4365/ad0a6c},
archivePrefix = {arXiv},
       eprint = {2303.05446},
 primaryClass = {astro-ph.SR},
       adsurl = {https://ui.adsabs.harvard.edu/abs/2024ApJS..270...15A}
}

@article{chatterjee2,
  author  = {Chatterjee, Subhamoy and Dayeh, Maher A. and Mu{\~n}oz-Jaramillo, Andr{\'e}s and Bain, Hazel M. and Moreland, Kimberly and Hart, Samuel},
  title   = {{MEMPSEP-I}. Forecasting the Probability of Solar Energetic Particle Event Occurrence Using a Multivariate Ensemble of Convolutional Neural Networks},
  journal = {Space Weather},
  year    = {2024},
  volume  = {22},
  number  = {9},
  pages   = {e2023SW003568},
  doi     = {10.1029/2023SW003568},
  eprint  = {2309.14570},
  archivePrefix = {arXiv}
}

@article{reames2004,
title = {Solar energetic particle variations},
journal = {Advances in Space Research},
volume = {34},
number = {2},
pages = {381-390},
year = {2004},
note = {Solar Variability and Climate Change},
issn = {0273-1177},
doi = {https://doi.org/10.1016/j.asr.2003.02.046},
url = {https://www.sciencedirect.com/science/article/pii/S0273117704002406},
author = {D.V Reames}}

@article{Visscher2015,
  author    = {De Visscher, R. and Delouille, V. and Dupont, P. and Deledalle, C.-A.},
  title     = {Supervised classification of solar features using prior information},
  journal   = {Journal of Space Weather and Space Climate},
  year      = {2015},
  volume    = {5},
  pages     = {A34},
  doi       = {10.1051/swsc/2015033}
}

@article{Muller-Mellin1995,
	author = {Müller-Mellin, R. and others},
	title = {{COSTEP} - Comprehensive Suprathermal and Energetic Particle Analyzer},
	journal = {Solar Physics},
	year = {1995},
	volume = {162},
	pages = {483-504},
	doi = {10.1007/BF00733437}
}

@article{Gold1998,
	author = {Gold, R. E. and others},
	title = {Electron, Proton, and Alpha Monitor on the Advanced Composition Explorer},
	journal = {Space Science Reviews},
	year = {1998},
	volume = {86},
	pages = {541-562},
	doi = {10.1023/A:1005088123348}
}

@article{sepnet,
  author  = {Yu, Yian and Chen, Yang and Zhao, Lulu and Whitman, Kathryn and Manchester, Ward and Gombosi, Tamas},
  title   = {Solar Energetic Particle Forecasting with Multi-Task Deep Learning: SEPNET},
  journal = {arXiv},
  year    = {2026},
  doi     = {10.48550/arXiv.2512.12786},
  url     = {https://arxiv.org/abs/2512.12786}
}

@article{Bloomfield2012,
  author  = {Bloomfield, D. S. and Higgins, P. A. and McAteer, R. T. J. and Gallagher, P. T.},
  title   = {Toward Reliable Benchmarking of Solar Flare Forecasting Methods},
  journal = {The Astrophysical Journal},
  year    = {2012},
  volume  = {747},
  number  = {2},
  pages   = {L41},
  doi     = {10.1088/2041-8205/747/2/L41}
}

@ARTICLE{Whitman2023,
       author = {{Whitman}, Kathryn and {Egeland}, Ricky and {Richardson}, Ian G. and {Allison}, Clayton and {Quinn}, Philip and {Barzilla}, Janet and {Kitiashvili}, Irina and {Sadykov}, Viacheslav and {Bain}, Hazel M. and {Dierckxsens}, Mark and {Mays}, M. Leila and {Tadesse}, Tilaye and {Lee}, Kerry T. and {Semones}, Edward and {Luhmann}, Janet G. and {N{\'u}{\~n}ez}, Marlon and {White}, Stephen M. and {Kahler}, Stephen W. and {Ling}, Alan G. and {Smart}, Don F. and {Shea}, Margaret A. and {Tenishev}, Valeriy and {Boubrahimi}, Soukaina F. and {Aydin}, Berkay and {Martens}, Petrus and {Angryk}, Rafal and {Marsh}, Michael S. and {Dalla}, Silvia and {Crosby}, Norma and {Schwadron}, Nathan A. and {Kozarev}, Kamen and {Gorby}, Matthew and {Young}, Matthew A. and {Laurenza}, Monica and {Cliver}, Edward W. and {Alberti}, Tommaso and {Stumpo}, Mirko and {Benella}, Simone and {Papaioannou}, Athanasios and {Anastasiadis}, Anastasios and {Sandberg}, Ingmar and {Georgoulis}, Manolis K. and {Ji}, Anli and {Kempton}, Dustin and {Pandey}, Chetraj and {Li}, Gang and {Hu}, Junxiang and {Zank}, Gary P. and {Lavasa}, Eleni and {Giannopoulos}, Giorgos and {Falconer}, David and {Kadadi}, Yash and {Fernandes}, Ian and {Dayeh}, Maher A. and {Mu{\~n}oz-Jaramillo}, Andr{\'e}s and {Chatterjee}, Subhamoy and {Moreland}, Kimberly D. and {Sokolov}, Igor V. and {Roussev}, Ilia I. and {Taktakishvili}, Aleksandre and {Effenberger}, Frederic and {Gombosi}, Tamas and {Huang}, Zhenguang and {Zhao}, Lulu and {Wijsen}, Nicolas and {Aran}, Angels and {Poedts}, Stefaan and {Kouloumvakos}, Athanasios and {Paassilta}, Miikka and {Vainio}, Rami and {Belov}, Anatoly and {Eroshenko}, Eugenia A. and {Abunina}, Maria A. and {Abunin}, Artem A. and {Balch}, Christopher C. and {Malandraki}, Olga and {Karavolos}, Michalis and {Heber}, Bernd and {Labrenz}, Johannes and {K{\"u}hl}, Patrick and {Kosovichev}, Alexander G. and {Oria}, Vincent and {Nita}, Gelu M. and {Illarionov}, Egor and {O'Keefe}, Patrick M. and {Jiang}, Yucheng and {Fereira}, Sheldon H. and {Ali}, Aatiya and {Paouris}, Evangelos and {Aminalragia-Giamini}, Sigiava and {Jiggens}, Piers and {Jin}, Meng and {Lee}, Christina O. and {Palmerio}, Erika and {Bruno}, Alessandro and {Kasapis}, Spiridon and {Wang}, Xiantong and {Chen}, Yang and {Sanahuja}, Blai and {Lario}, David and {Jacobs}, Carla and {Strauss}, Du Toit and {Steyn}, Ruhann and {van den Berg}, Jabus and {Swalwell}, Bill and {Waterfall}, Charlotte and {Nedal}, Mohamed and {Miteva}, Rositsa and {Dechev}, Momchil and {Zucca}, Pietro and {Engell}, Alec and {Maze}, Brianna and {Farmer}, Harold and {Kerber}, Thuha and {Barnett}, Ben and {Loomis}, Jeremy and {Grey}, Nathan and {Thompson}, Barbara J. and {Linker}, Jon A. and {Caplan}, Ronald M. and {Downs}, Cooper and {T{\"o}r{\"o}k}, Tibor and {Lionello}, Roberto and {Titov}, Viacheslav and {Zhang}, Ming and {Hosseinzadeh}, Pouya},
        title = "{Review of Solar Energetic Particle Prediction Models}",
      journal = {Advances in Space Research},
         year = 2023,
        month = dec,
       volume = {72},
       number = {12},
        pages = {5161-5242},
          doi = {10.1016/j.asr.2022.08.006},
       adsurl = {https://ui.adsabs.harvard.edu/abs/2023AdSpR..72.5161W}
}

@ARTICLE{Kasapis2026,
       author = {{Kasapis}, Spiridon and {Hosseinzadeh}, Pouya and {Whitman}, Kathryn and {Egeland}, Ricky and {Georgoulis}, Manolis and {Vourlidas}, Angelos and {Papaioannou}, Athanasios and {Lavasa}, Eleni and {Anastasiadis}, Anastasios and {Giannopoulos}, Giorgos and {Munoz-Jaramillo}, Andres and {Poduval}, Bala and {Kitiashvili}, Irina N. and {Kosovichev}, Alexander G. and {Sadykov}, Viacheslav and {Filali Boubrahimi}, Soukaina and {Hutchins}, Tate T. and {Farooki}, Hameedullah A. and {Cuesta}, Manuel E. and {Khoo}, Leng Y. and {Pak}, Sungmin and {Czarnota}, Robert and {Rankin}, Jamie S. and {Szalay}, Jamey and {Shen}, Mitchell M. and {Livadiotis}, Georgios and {Xu}, Zigong and {McComas}, David J. and {Sarlis}, Nikolaos and {Hristopulos}, Dionissios and {Posner}, Arik and {Engell}, Alec J. and {AbuBakr Ali}, Mohammed and {Abdelkawy}, Ali G.~A. and {Shaltout}, Abdelrazek M.~K. and {Beheary}, M.~M. and {Lee}, Christina O. and {Aminalragia-Giamini}, Sigiava and {Papadimitriou}, Constantinos and {Sandberg}, Ingmar and {Raptis}, Savvas and {Hamdi}, Shah Muhammad and {Laurenza}, Monica and {Stumpo}, Mirko and {Rotti}, Sumanth A. and {Jackson}, India and {Ali}, Aatiya and {Gunes Baydin}, Atilim and {Schwadron}, Nathan and {Chatterjee}, Subhamoy and {Dayeh}, Maher A. and {Nita}, Gelu M. and {O'Keefe}, Patrick M. and {Jie Chong}, Chun and {Kosovich}, Paul and {Marroquin}, Russell D. and {Aydin}, Berkay and {Martens}, Petrus C. and {Zhao}, Lulu and {Chen}, Yang and {Yu}, Yian and {Bobra}, Monica G. and {Manchester}, Ward and {Gombosi}, Tamas and {Zhang}, Ming and {Torres}, Jesse and {Chan}, Philip K. and {Nedal}, Mohamed and {Kozarev}, Kamen and {Zhang}, Peijin and {Moreland}, Kimberly and {Bain}, Hazel M. and {Hart}, Samuel and {Starkey}, Michael J. and {Ling}, Alan G. and {Benella}, Simone},
        title = "{Review of Machine Learning Models for Solar Energetic Particle Prediction}",
      journal = {arXiv e-prints},
         year = 2026,
        month = jun,
          eid = {arXiv:2606.19539},
        pages = {arXiv:2606.19539},
          doi = {10.48550/arXiv.2606.19539},
archivePrefix = {arXiv},
       eprint = {2606.19539},
 primaryClass = {astro-ph.SR},
       adsurl = {https://ui.adsabs.harvard.edu/abs/2026arXiv260619539K}
}

@article{Blunier2021,
  author    = {Blunier, S. and Toledo, B. and Rogan, J. and Valdivia, J. A.},
  title     = {A Nonlinear System Science Approach to Find the Robust Solar Wind Drivers of the Multivariate Magnetosphere},
  journal   = {Space Weather},
  year      = {2021},
  volume    = {19},
  number    = {6},
  pages     = {e2020SW002634},
  doi       = {10.1029/2020SW002634}
}

@ARTICLE{Ali2026JGRA_GOES_EPHIN,
       author = {{Ali}, Aatiya and {Sadykov}, Viacheslav},
        title = "{Comparative Analysis of 10 - 50 MeV Solar Proton Events at Lagrange Point 1 and the Geostationary Orbit}",
      journal = {Journal of Geophysical Research (Space Physics)},
         year = 2026,
        month = jun,
       volume = {131},
       number = {6},
          eid = {e2025JA034853},
        pages = {e2025JA034853},
          doi = {10.1029/2025JA034853},
archivePrefix = {arXiv},
       eprint = {2511.04793},
 primaryClass = {astro-ph.SR},
       adsurl = {https://ui.adsabs.harvard.edu/abs/2026JGRA..13134853A}
}

@INPROCEEDINGS{Ali11415910_MLcontamination,
  author={Ali, Aatiya and Sadykov, Viacheslav},
  booktitle={2025 IEEE International Conference on Data Mining Workshops (ICDMW)}, 
  title={Using ML-Based Regression Techniques to Mitigate Goes Energetic Proton Flux Data Contamination and Magnetospheric Effects}, 
  year={2025},
  volume={},
  number={},
  pages={1923-1930},
  doi={10.1109/ICDMW69685.2025.00231}}

@inproceedings{paszke2019pytorch,
    author        = {Paszke, Adam and Gross, Sam and Massa, Francisco and
                     Lerer, Adam and Bradbury, James and Chanan, Gregory and
                     Killeen, Trevor and Lin, Zeming and Gimelshein, Natalia and
                     Antiga, Luca and Desmaison, Alban and K{\"o}pf, Andreas and
                     Yang, Edward and DeVito, Zachary and Raison, Martin and
                     Tejani, Alykhan and Chilamkurthy, Sasank and Steiner, Benoit and
                     Fang, Lu and Bai, Junjie and Chintala, Soumith},
    title         = {{PyTorch}: An Imperative Style, High-Performance Deep Learning Library},
    booktitle     = {Advances in Neural Information Processing Systems},
    volume        = {32},
    pages         = {8024--8035},
    year          = {2019},
    eprint        = {1912.01703},
    archivePrefix = {arXiv},
    primaryClass  = {cs.LG},
    doi           = {10.48550/arXiv.1912.01703}
}
\bibliographystyle{aasjournalv7}
\end{document}